\documentclass[11pt]{article}

\input{epsf}
\usepackage{epsfig}
\usepackage{amssymb}
\usepackage{amsfonts}
\usepackage{amsbsy}
\usepackage[all]{xy}
\usepackage{amsmath}
\usepackage{esint}
\usepackage{xcolor}
\definecolor{burgundy}{rgb}{0.5, 0.0, 0.13}
\usepackage[linktocpage=true,colorlinks=true,linkcolor=burgundy,citecolor=black!20!blue,urlcolor=violet]{hyperref}

\usepackage{amssymb,amscd}
\usepackage{mathrsfs}
\usepackage{amsmath,amsthm}
\usepackage{dsfont}
\usepackage{soul}
\usepackage{comment}
\usepackage{cite} 

\def\p{\partial}

\def\CD {{\cal D}}

\def\CI {{\cal I}}

\def\CL {{\cal L}}

\def\CN {{\cal N}}

\def\CR {{\cal R}}

\def\CI {{{\cal I}}}

\def\CT {{\cal T}}

\def\IC{\mathbb{C}}

\def\IP{\mathbb{P}}
\def\IQ{\mathbb{Q}} 
\def\IR{{\mathbb{R}}}

\def\IZ{{\mathbb{Z}}}

\def\ff{\mathfrak{f}}

\def\fg{\mathfrak{g}}

\def\fl{\mathfrak{l}}

\def\fu{\mathfrak{u}}

\def\fB{\mathfrak{B}}

\def\fQ{\mathfrak{Q}}

\def\Dslash{\,{\raise.15ex\hbox{/}\mkern-12mu \CD}}

\def\rmk#1{\bigskip\noindent{\bf Remarks} }

\usepackage{tikz}
\usepackage{tikz-cd}
\usetikzlibrary{arrows}
\usetikzlibrary{arrows.meta}
\usetikzlibrary{positioning}
\usetikzlibrary{shapes,snakes}
\usetikzlibrary{fit}
\usetikzlibrary{decorations.pathmorphing,decorations.pathreplacing,decorations.markings}

\def\lm{\limits}

\def\I{{\rm i}}

\newcommand\sqbox[1]{{
		\setbox0=\hbox{\mbox{$\Box$}}
		\setbox1=\hbox{\mbox{\raisebox{0.35ex}{\tiny #1}}}
		\mbox{\raisebox{-0.2ex}{\rlap{\hbox to \wd0{\hss{\box1}\hss}}\box0}}
}}

\newcommand\Kappa{\mathrm{K}}

\def\myblue{white!40!blue}

\hoffset= - 0.9in

\numberwithin{equation}{section}
\numberwithin{theorem}{section}

\begin{document}

\pagenumbering{Alph} 
\begin{titlepage}
	
	\begin{center}
		\hfill\break
		
		\vspace{2.5cm}
		
		{\bf\Large{BPS States Meet Generalized Cohomology}}
		\vskip 1cm 
		\renewcommand{\thefootnote}{}
		{Dmitry Galakhov\footnote[2]{e-mail: d.galakhov.pion@gmail.com; galakhov@itep.ru}} 
		\vskip 0.2in 
		
		\renewcommand{\thefootnote}{\roman{footnote}}
		
		{\small{ 
				\textit{Institute for Information Transmission Problems,\\}
				\textit{ Moscow, 127051, Russia,\\}
				\textit{NRC ``Kurchatov Institute'',\\}
				\textit{Moscow, 123182, Russia}
		}}
	
	\end{center}
	
	\vskip 0.5in
	\baselineskip 16pt
	
	\begin{abstract}
		In this note we review a construction of a BPS Hilbert space in an effective supersymmetric quiver theory with 4 supercharges.
		We argue abstractly that this space contains elements of an equivariant generalized cohomology theory $E_G^{*}(-)$ of the quiver representation moduli space giving concretely Dolbeault cohomology, K-theory or elliptic cohomology depending on the spacial slice is compactified to a point, a circle or a torus respectively, and something more amorphous in other cases.
		Furthermore BPS instantons -- basic contributors to interface defects or a Berry connection -- induce a \emph{BPS algebra} on the BPS Hilbert spaces representing Fourier-Mukai transforms on the quiver representation moduli spaces descending to an algebra over $E_G^{*}(-)$ as its representation.
		In the cases when the quiver describes a toric Calabi-Yau three-fold (CY${}_3$) the algebra is a respective generalization of the quiver BPS Yangian algebra discussed in the literature, in more general cases it is given by an abstract generalized cohomological Hall algebra.

	\end{abstract}
	
	\date{\today}
\end{titlepage}
\pagenumbering{arabic} 

\newpage
\setcounter{tocdepth}{2}
\tableofcontents

\section{Summary and discussion}

Symmetry acquires a cornerstone position in a discussion of any physical system.
It is not surprising that recent developments (see reviews \cite{Cordova:2022ruw, Bah:2022wot} and references therein) in generalizing the notion of symmetry attracts a lot of physics community attention.
It is peculiar that this novel notion incorporates non-perturbative aspects and fascinating properties of quantum field theory defects, quasi-particles and other non-local objects.
In this paper we will turn to a modest manifestation of this broad concept and concentrate on BPS algebras emerging in considerations of BPS states in supersymmetric theories.

The notion of the BPS algebra goes back to \cite{Harvey:1995fq}, and has been developed intensely since (see \cite{Harrison:2021gnp, Rapcak:2021hdh} and references therein for recent reviews).
We will further nail down our interest mostly to a specific family of BPS algebras known as quiver Yangians \cite{Li:2020rij, Bao:2022jhy} and their generalizations (see e.g. \cite{Noshita:2021ldl}).
The quiver Yangian algebra emerges in an effective description of a D-brane system on a toric Calabi-Yau three-fold (CY${}_3$) as a quiver supersymmetric quantum mechanics with 4 supercharges -- a compactification of a 4d $\CN=1$ Yang-Mills-Higgs theory.
In \cite{Galakhov:2021vbo} it was shown that toroidal and elliptic quiver Yangian generalizations could be constructed as BPS algebras of states in the 4d $\CN=1$ quiver Yang-Mills-Higgs theory compactified on a circle or a torus.
The BPS algebra acts on the BPS states, and its physical construction allows one to calculate defining relations between matrix elements of the  prospective algebra.
Crucial ingredients of those relations are one-loop determinants emerging due to localization and falling into a systematization analogous to the rational/trigonometric/elliptic hierarchy for the algebras in question.
In the case of the determinant its building block -- one-loop determinant of the Higgs chiral supermultiplet -- is given by a linear/trigonometric/elliptic function suggesting that the respective one-loop determinant is a generalized genus corresponding to cohomology/K-theory/elliptic cohomology of the target space:
\begin{equation}\label{main_table}
	\begin{array}{c|c|c|c}
		\mbox{Chern class} & \mbox{Algebra} & \mbox{Cohomological theory} & \mbox{QFT space}\\
		\hline
		z & \mbox{Affine Yangian} & \mbox{Dolbeault cohomology} & \mbox{point}\\
		\sinh \beta z & \mbox{DIM (toroidal)} & \mbox{K-theory} & S^1\\
		\vartheta_{11}(z|\tau) & \mbox{Elliptic} & \mbox{Elliptic cohomology} & T^2\\
		{\bf e}_{\xi}(z) & \mbox{\it Does it exist?} & \mbox{Algebraic cobordism }\Omega_*^{U} & \Sigma_g?
	\end{array}
\end{equation}
Arranged in such a table data suggest an extension of this story to a generalized cohomology theory (GCT) of algebraic cobordisms \cite{levine2001cobordisme}.
The role of the one-loop determinant for GCT is played by an ``exponent'' function ${\bf e}_{\xi}(z)$ -- the inverse function of the formal group law logarithm \cite{Quillen}:
\begin{equation}
	\log_{\xi}z:=z+\frac{\xi[\IC\IP^1]}{2}z^{2}+\frac{\xi[\IC\IP^2]}{3}z^{3}+\frac{\xi[\IC\IP^3]}{4}z^{4}+\frac{\xi[\IC\IP^4]}{5}z^{5}+\ldots\,,
\end{equation}
where $\xi:\;\Omega_*^{U}\longrightarrow \Lambda$ is a homomorphism of an algebraic cobordism ring to a Lazard ring.

The major question we would like to pose in this note is if based on hints of \eqref{main_table} the QFT provides a basis for generalized cohomology as it does in the case of the Morse theory \cite{witten1982supersymmetry, Bullimore:2016hdc} or its categorical lifts \cite{Gaiotto:2015aoa, Dedushenko:2021mds, Bullimore:2021rnr, Galakhov:2020upa, Chen:2020iyo, Khan:2020hir, Dimofte:2019zzj}, at least for the mentioned family of quiver models.
For a recent discussion on similar questions for elliptic cohomology and BPS states in 3d $\CN=4$ theories see \cite{Bullimore:2021auw,Dedushenko:2021mds,Bullimore:2021rnr,Dedushenko:2022pem}.
We are inclined to give a positive answer, however let us raise first some problems accompanying this construction.

The QFT construction for the BPS algebra describes its representation rather than the algebra itself.
The BPS Hilbert space is a representation vector space where the algebra acts, and matrix coefficients for certain QFT operators reflect some matrix coefficients of the algebra in the given representation.
Such a generalized matrix representation could be constructed on its own without a reference to its algebraic structure.
For instance, in \cite{Awata:2020xfq} a generalization was constructed for the Macdonald polynomial representation of the DIM algebra -- associated with affine Dynkin diagram of $\widehat{\fg\fl}_1$, or a simple toric Calabi-Yau $\IC^3$ -- in terms of generic functions ${\bf e}_{\xi}(z)$.
However the question if this matrix algebra could be refurbished in the language of a universal enveloping algebra as a ring of generators modulo some set of Poincar\'e-Birkhoff-Witt relations is open.

Another question we will intentionally avoid discussing in great detail in this paper is if a twisted compactification on a higher genus Riemann surface $\Sigma_g$ \cite{Bullimore:2021auw, Ferrari:2022opl} is an exhaustive family of descriptions for GCTs as a natural extension of table \eqref{main_table} in the last column suggests.
Instead we argue that the essential information about the BPS Hilbert space (its flux-less subsector), or its GCT avatar, behavior can be packed in a universal function ${\bf e}_{\xi}(z)$ without an explicit reference to  $\Sigma_g$ or $\xi$.
This idea is somewhat reminiscent of the construction for topological modular forms \cite{Lurie, hopkins1995topological}.
Similarly to the universal dependence of Witten's genera on the modular parameter we realize that the QFT answers for our model families are universal in some class of functions and suspect a generalized cohomology theory is responsible for this universality.
It would be interesting to investigate further if our construction is helpful in the quest to produce QFT constructions of topological modular forms \cite{stolz2011supersymmetric, Gaiotto:2019gef}.

Also we would like to translate into the current framework an idea of \cite{Gaiotto:2015aoa, Brunner:2008fa, Carqueville:2016kdq, Kong:2022cpy}: interface defects providing Berry connections on Hilbert spaces induce morphisms on geometric structures emerging in the physical systems.
For instance, soliton amplitudes in linear gauged sigma-models support Fourier-Mukai transforms on the derived coherent sheave category of D-branes \cite{Galakhov:2020upa} .
In particular, we construct matrix coefficients for the resulting BPS algebra as elements of instanton amplitudes, also in the universal language of ${\bf e}_{\xi}(z)$.
In a similar fashion instantons in the effective quiver QFTs support Fourier-Mukai transforms of quiver varieties associated with Hecke modifications of D-brane sheaves on CY${}_3$.
We demonstrate that relations between matrix generators of the generalized BPS algebra may be treated as monodromy relations induced by an effective Berry connection.

It is worth pointing out that our discussion of GCTs is somewhat perpendicular to the usual applications of generalized (co)homology and (co)bordism theories in the physics of anomalies. Unfortunately, it is beyond our scope to review the vast literature on this broad subject.
The interested reader might consult with recent sources on this topic such as \cite{Wan:2018bns, Cordova:2022ruw}.
One crucial catching eye difference is that the homological degree in our case is not directly related to a dimension of the target space or of the world-sheet of the theory, rather it is given by the R-charge, or by the fermion number as in the canonical setting of  \cite{witten1982supersymmetry}.
However our case seems not to fall out a TQFT/cobordism correspondence hypothesis \cite{Baez:1995xq,Freed:2012hx,2009arXiv0905.0465L}.
For an anomaly from anomaly inflow one would expect a TQFT with a Schwarz-type topological action (when the action formulation is tractable).
Whereas our construction relies heavily on supersymmetry of the theory, and it would be natural to conjecture that the bare BPS spectrum decoupled from the non-BPS states has some IR Witten-type TQFT description.
On the other hand it is also worth mentioning that this decoupling in the construction of the BPS algebra, for example, is rather artificial.
In practice, BPS and non-BPS states get mixed as one intersects marginal stability walls on the moduli space due to wall-crossing phenomena.

Our construction is naturally applicable to the effective quiver description of D-brane systems on toric CY${}_3$'s. 
Therefore it would be natural to ask if some elements of this construction could be transferred to a discussion of D-brane systems on toric CY${}_4$'s \cite{Nekrasov:2018xsb, Franco:2019bmx, Cao:2019tvv, Bae:2022pif, Szabo:2023ixw}, if similar manipulations with instantons and defects allow one to produce some novel BPS algebra.

This paper is organized as follows.
In section \ref{s:preliminaries} we discuss some preliminary definitions in quiver QFTs we will apply throughout the paper.
Section \ref{s:GCT} is devoted to a mathematical side of the discussion: we briefly review the cobordism generalized cohomology theory and localization due to the equivariant symmetry action, discuss some aspects of the mathematical BPS algebra constructions.
In section \ref{s:phys_BPS} we argue that the BPS Hilbert space of theories in question deliver a description for a generalized cohomology theory of the target space and calculate relations between the BPS algebra matrix coefficients via instanton amplitudes.
In the appendix we placed some auxiliary information on the Berry connection in a simple fermion system.


\section{Preliminaries on quiver varieties}\label{s:preliminaries}


\subsection{Effective D-brane description}

It was shown in the classic paper \cite{Douglas:1996sw} that IR dynamics of D-brane systems wrapping holomorphic cycles of Calabi-Yau manifolds is described nicely by quiver QFTs.
Some details on various constructions allowing one to associate a D-brane system to a quiver theory could be found in many literature sources \cite{Denef:2002ru, Aganagic:2003db, Ooguri:2009ijd, Yamazaki:2008bt, Li:2020rij, Hanany:2005ss, Szendroi:2007nu}, see also references therein.
Let us mention that a family of toric Calabi-Yau three-folds gives rise to a family of associated BPS algebras -- quiver Yangians \cite{Li:2020rij}.
In this paper we will not dive into details of this Calabi-Yau -- quiver construction, rather we concentrate on quiver field theories defined by the following data: a quiver and a superpotential $(\fQ,W)$.

Starting from these data an effective theory can be constructed.
This theory appears as a dimensional reduction of the 4d $\CN=1$ super-Yang-Mills-Higgs theory with the standard Lagrangian whose matter-potential content is defined by $(\fQ,W)$.

Here we are planning to adopt some common notations for purposes we are planning to pursue throughout the paper.

The set of quiver nodes we denote as $\fQ_0$.
To each node one associates a gauge group $U(n)$ for some $n$ and corresponding vector multiplet:
\begin{equation}
	A_{\mu},\quad \lambda_{\alpha},\quad { \bf D}\,.
\end{equation}
We would like to rename and reorganize components of the 4d vector potential in the following way:
\begin{equation}
	A_{\mu}=(A_0,A_1=\sigma+\bar\sigma,A_2=\I(\sigma-\bar\sigma),A_3=h)\,,
\end{equation}
so that:
\begin{equation}
	\begin{split}
	D_z=\frac{1}{2}(D_1-\I D_2)=\p_z+\I\sigma,\; D_{\bar z}=\frac{1}{2}(D_1+\I D_2)=\p_{\bar z}+\I\bar\sigma,\; A_1dx^1+A_2dx^2=\sigma dz+\bar\sigma d\bar z\,.
\end{split}
\end{equation}

The set of arrows we denote as $\fQ_1$.
To arrows one associates chiral multiplets:
\begin{equation}
	\phi,\quad \psi_{\alpha},\quad {\bf F}\,.
\end{equation}
This field is charged as the fundamental rep $n$  with respect to corresponding $U(n)$ put in the head node of the arrow, and with respect to the anti-fundamental rep $\bar m$ under corresponding arrow tail node $U(m)$.

Let us demonstrate here an explicit form of the supercharges in a dimensionally reduced $\CN=4$ SQM corresponding to a single chiral field charged with respect to a single $U(n)$ following \cite{Wess:1992cp}. There is no problem to generalize these expressions to a generic quiver.
\begin{equation}\label{Qs}
	\begin{split}
		Q_1={\rm Tr}\,\Big[ & \bar\lambda_{\dot 2}\left(\p_h+2\left[\sigma,\bar\sigma\right]+{\bf D}\right)-\bar\lambda_{\dot 1}\left(\p_{\bar \sigma}-2\left[\sigma,h\right]\right)-\\
		& -\sqrt{2}\I\psi_1\left(\p_{\phi}-\bar\phi h\right)+2\sqrt{2}\I \psi_2\bar\phi\sigma-\sqrt{2}\I\bar\psi_{\dot 2}{\bf F}\Big]\,,\\
		Q_2={\rm Tr}\,\Big[ & \bar\lambda_{\dot 1}\left(\p_h+2\left[\sigma,\bar\sigma\right]-{\bf D}\right)+\bar\lambda_{\dot 2}\left(\p_{\sigma}+2\left[\bar\sigma,h\right]\right)-\\
		& -\sqrt{2}\I\psi_2\left(\p_{\phi}+\bar\phi h\right)+2\sqrt{2}\I \psi_1\bar\phi\bar\sigma+\sqrt{2}\I\bar\psi_{\dot 1}{\bf F}\Big]\,,\\
		\bar Q_{\dot 1}={\rm Tr}\,\Big[ & -\lambda_{2}\left(\p_h-2\left[\sigma,\bar\sigma\right]-{\bf D}\right)+\lambda_{1}\left(\p_{\sigma}-2\left[\bar\sigma,h\right]\right)-\\
		& -\sqrt{2}\I\bar\psi_{\dot 1}\left(\p_{\bar\phi}+h\phi\right)-2\sqrt{2}\I \psi_2\bar\sigma\phi+\sqrt{2}\I\psi_{2}{\bf F}^{\dagger}\Big]\,,\\
		\bar Q_{\dot 2}={\rm Tr}\,\Big[ & -\lambda_{1}\left(\p_h-2\left[\sigma,\bar\sigma\right]+{\bf D}\right)-\lambda_{2}\left(\p_{\bar \sigma}+2\left[\sigma,h\right]\right)-\\
		& -\sqrt{2}\I\bar\psi_{\dot 2}\left(\p_{\bar\phi}-h\phi\right)-2\sqrt{2}\I \bar\psi_{\dot 1}\sigma\phi-\sqrt{2}\I\psi_{1}{\bf F}^{\dagger}\Big]\,.
	\end{split}
\end{equation}
Here $\bf D$ and $\bf F$ are expectation values for the auxiliary fields.
For a generic quiver we could write explicit expressions \cite{Alim:2011kw}:
\begin{equation}
	\begin{split}
	{\bf D}_i=r_i-\sum\lm_{(a:*\to i)\in\fQ_1}\phi_a\phi_a^{\dagger}+\sum\lm_{(b:i\to *)\in\fQ_1}\phi_b^{\dagger}\phi_b,\quad \forall i\in \fQ_0\,;\\
	{\bf F}_a=-\overline{\p_{\phi_a}W},\quad {\bf F}^{\dagger}_a=-\p_{\phi_a}W,\quad \forall a\in\fQ_1\,,
	\end{split}
\end{equation}
where $r_i$ are FI parameters of the theory.

If the theory has angular rotation isometry, say, one has compactified the space to a point, or there is an $\Omega$-background along a plane in 3d space, one can introduce a generator of these angular rotations $J_3$ commuting with the superchages in a specific way:
\begin{equation}
	\left[J_3,Q_1\right] =-\frac{1}{2}Q_1, \quad \left[J_3,Q_2\right] =\frac{1}{2}Q_2,\quad 	\left[J_3,\bar Q_{\dot 1}\right] =\frac{1}{2}\bar Q_{\dot 1}, \quad \left[J_3,\bar Q_{\dot 2}\right] =-\frac{1}{2}\bar Q_{\dot 2}\,.
\end{equation}
This operator could be combined with the R-charge generator:
\begin{equation}
	\left[R,Q_{\alpha}\right]=\frac{1}{2}Q_{\alpha},\quad \left[R,\bar Q_{\dot\alpha}\right]=\frac{1}{2}\bar Q_{\dot\alpha}\,,
\end{equation}
so that the resulting isospin generator $I_3^{\pm}=J_3\pm R$ commutes with either $Q_1$, $\bar Q_{\dot 1}$ or $Q_2$, $\bar Q_{\dot 2}$ respectively.
The isospin generator could be used to produce a refined spin index \cite{Manschot:2010qz, Manschot:2011xc,Gaiotto:2010okc}.
Respectively we could choose either supercharge generator to construct the Hilbert subspace of BPS states.
Without loss of generality we choose $Q_1$ and construct the BPS Hilbert space as cohomology of the target space \cite{witten1982supersymmetry,Bullimore:2016hdc}:
\begin{equation}\label{H_BPS}
	\mathscr{H}_{\rm BPS}\cong H^*\left({\rm Rep}(\fQ),Q_1\right)\,.
\end{equation}
The resulting BPS Hilbert space is spanned by states annihilated by all the 4 supercharges \cite{Denef:2002ru}.
However it is useful to select one for more explicit localization as certain holomorphic structures become fixed.


\subsection{Higgs-Coulomb duality}\label{sec:HC_du}

The chosen supercharge $Q_1$ has a form of a deRahm-Dolbeault equivariant differential twisted with a superpotential \cite{Galakhov:2020vyb}:
\begin{equation}\label{supercharge}
	Q_1=e^{-U}\left(d_{h}+\p_{\bar\sigma,\phi}+\iota_{\bar V}+d\bar W\right)e^{U}\,,
\end{equation}
where the Morse height function reads:
\begin{equation}\label{height}
	U=\sum\lm_{i\in\fQ_0}{\rm Tr}\,h_i\left(r_i-\sum\lm_{(a:*\to i)\in\fQ_1}\phi_a\phi_a^{\dagger}+\sum\lm_{(b:i\to *)\in\fQ_1}\phi_b^{\dagger}\phi_b\right)\,,
\end{equation}
and the vector field is induced by a holomorphic complexified action of the gauge and flavor groups:
\begin{equation}
	\bar V=\sum\lm_{(a:i\to j)\in \fQ_1}\left(\phi_a^{\dagger}\sigma_j-\sigma_i\phi_a^{\dagger}-\mu_a\phi_a^{\dagger}\right)\frac{\p}{\p\phi_a^{\dagger}}\,,
\end{equation}
where $\mu_a$ is a flavor charge -- complex mass -- of field $\phi_a$.

The localization procedure allows one to approximate the BPS Hilbert space by the classical vacuum wave functions defined by critical points of the height function and superpotential fixed with respect to the action of the vector filed.
In practice, the effective energy potential is given by:
\begin{equation}\label{energy}
	E\sim |\nabla U|^2+|\nabla W|^2+|V|^2\,.
\end{equation}

A canonical formulation of the Higgs-Coulomb duality \cite{Denef:2002ru} in a quiver gauge theory is rather transparent from this point of view.
It follows naturally from the fact that it is impossible to assign non-trivial vacuum expectation values to scalars $\sigma_i$ in the gauge multiplet (if $\mu_a=0$) and to scalars $\phi_a$ in the chiral multiplet so that they solve the vacum equations simultaneously.
The localization procedure allows one to choose a direction in the parameter space towards the IR regime.
As a result of this choice only a half of the vacuum equations for vevs are taken into account at the zeroth order of approximation. 
The rest is considered in higher orders and acquires loop corrections -- therefore only one ``dominant" type of fields, $\sigma_i$'s or $\phi_a$'s, acquires vevs.
However two inequivalent direction choices in the parameter space should produce equivalent localization pictures inducing an equivalence, a duality, an isomorphism of the BPS Hilbert spaces in two quite different languages.

In the case of non-zero masses $\mu_a$ the Higgs localization branch turns out to be a mixed one since zero vevs get resolved by $\mu_a$ values.
We would ignore this terminological fact and keep calling it a Higgs branch since the effective theory has the quiver representation variety parameterized by $\phi_a$ as a target space, and $\mu_a$'s parameterize an  equivariant isometry action on it.

Further this section we would like to describe schematically specifics of wave functions on both branches.

\begin{figure}[h!]
	\begin{center}
		\begin{tikzpicture}
			\node(A) at (0,0) {$
				\begin{array}{c}
					\begin{tikzpicture}[scale=0.7]
						\tikzset{ln/.style={thick}}
						\tikzset{blb/.style={fill=\myblue}}
						\begin{scope}[shift={(1.55422,0)}]
							\begin{scope}[rotate=-162]
								\begin{scope}[shift={(0,0.850651)}]
									\draw[ln] (0,0) -- (0,0.8) -- (0,1.3) (0,0.8) -- (0.433013, 1.05) (0,0.8) -- (-0.433013, 1.05);
									\draw[blb] (0,0.8) circle (0.1);
									\foreach \x/\y in {0/1.3, 0.433013/1.05, -0.433013/1.05}
									{
										\draw[blb] (\x,\y) circle (0.06);
									}
								\end{scope}
							\end{scope}
							\draw[ln] (0.850651,0) -- (1.850651,0);
							\draw[blb] (1.850651,0) circle (0.1);
							\draw[ln] (0.850651, 0.) to[out=111, in=321] (0.262866, 0.809017) (0.850651, 0.) to[out=141, in=291] (0.262866, 0.809017) (0.262866, 0.809017) (0.262866, 0.809017) -- (-0.688191,	0.5) (-0.688191, -0.5) -- (0.262866, -0.809017) -- (0.850651, 0.);
							\foreach \x/\y in {0.850651/0., 0.262866/0.809017, 0.262866/-0.809017}
							{
								\draw[blb] (\x,\y) circle (0.1);
							}
						\end{scope}
						\foreach \s in {60, 180}
						{
							\begin{scope}[rotate=\s]
								\begin{scope}[shift={(0,1)}]
									\draw[ln] (0,0) to[out=75,in=285] (0,1) (0,0) to[out=105,in=255] (0,1);
									\draw[blb] (0,1) circle (0.1);
								\end{scope}
							\end{scope}
						}
						\foreach \s in {0, 120}
						{
							\begin{scope}[rotate=\s]
								\begin{scope}[shift={(0,1)}]
									\draw[ln] (0,0) -- (0,0.8) -- (0,1.3) (0,0.8) -- (0.433013, 1.05) (0,0.8) -- (-0.433013, 1.05);
									\draw[blb] (0,0.8) circle (0.1);
									\foreach \x/\y in {0/1.3, 0.433013/1.05, -0.433013/1.05}
									{
										\draw[blb] (\x,\y) circle (0.06);
									}
								\end{scope}
							\end{scope}
						}
						\draw[ln] (0.866025, 0.5) -- (0., 1.) -- (-0.866025,0.5) -- (-0.866025, -0.5) -- (0., -1.) -- (0.866025, -0.5) (0.866025, -0.5) to[out=75,in=285] (0.866025, 0.5) (0.866025, -0.5) to[out=105,in=255] (0.866025, 0.5);
						\foreach \x/\y in {0.866025/0.5, 0./1., -0.866025/0.5, -0.866025/-0.5, 0./-1., 0.866025/-0.5}
						{
							\draw[blb] (\x,\y) circle (0.15);
						}
					\end{tikzpicture}
				\end{array}
				$};
			\node (B) at (-4,-2.5) {$
				\begin{array}{c}
					\begin{tikzpicture}[rotate=120]
						\begin{scope}[scale=0.7]
							\draw[thick, -stealth] (0,0) -- (0.866025, -0.5);
							\node[right] at (0.866025, -0.5) {$\scriptstyle {\rm Im}\,\sigma$};
							\draw[thick, -stealth] (0,0) -- (-0.866025, -0.5);
							\node[right] at (-0.866025, -0.5) {$\scriptstyle {\rm Re}\,\sigma$};
						\end{scope}
						\begin{scope}[shift={(0,0.3)}]
							\begin{scope}[scale=0.6]
								\begin{scope}[yscale=0.57735]
									\draw[fill=white!80!blue,postaction={decorate},
									decoration={markings, mark= at position 0.65 with {\arrow{stealth}}}] (0,0) circle (1);
									\draw[dashed, postaction={decorate},
									decoration={markings, mark= at position 0.65 with {\arrow{stealth}}}] (0,0) circle (0.7);
									\draw[dashed, postaction={decorate},
									decoration={markings, mark= at position 0.65 with {\arrow{stealth}}}] (0,0) circle (0.4);
									\draw[fill=black] (0,0) circle (0.05);
								\end{scope}
							\end{scope}
						\end{scope}
						\begin{scope}[shift={(0,0.8)}]
							\begin{scope}[scale=0.9]
								\begin{scope}[yscale=0.57735]
									\draw[fill=white!80!blue,postaction={decorate},
									decoration={markings, mark= at position 0.65 with {\arrow{stealth}}}] (0,0) circle (1);
									\draw[dashed, postaction={decorate},
									decoration={markings, mark= at position 0.65 with {\arrow{stealth}}}] (0,0) circle (0.7);
									\draw[dashed, postaction={decorate},
									decoration={markings, mark= at position 0.65 with {\arrow{stealth}}}] (0,0) circle (0.4);
									\draw[fill=black] (0,0) circle (0.05);
								\end{scope}
							\end{scope}
						\end{scope}
						\begin{scope}[shift={(0,1.3)}]
							\begin{scope}[scale=1.3]
								\begin{scope}[yscale=0.57735]
									\draw[fill=white!80!blue,postaction={decorate},
									decoration={markings, mark= at position 0.65 with {\arrow{stealth}}}] (0,0) circle (1);
									\draw[dashed, postaction={decorate},
									decoration={markings, mark= at position 0.65 with {\arrow{stealth}}}] (0,0) circle (0.7);
									\draw[dashed, postaction={decorate},
									decoration={markings, mark= at position 0.65 with {\arrow{stealth}}}] (0,0) circle (0.4);
									\draw[fill=black] (0,0) circle (0.05);
								\end{scope}
							\end{scope}
						\end{scope}
						\begin{scope}[shift={(0,1.7)}]
							\begin{scope}[scale=0.8]
								\begin{scope}[yscale=0.57735]
									\draw[fill=white!80!blue,postaction={decorate},
									decoration={markings, mark= at position 0.65 with {\arrow{stealth}}}] (0,0) circle (1);
									\draw[dashed, postaction={decorate},
									decoration={markings, mark= at position 0.65 with {\arrow{stealth}}}] (0,0) circle (0.7);
									\draw[dashed, postaction={decorate},
									decoration={markings, mark= at position 0.65 with {\arrow{stealth}}}] (0,0) circle (0.4);
									\draw[fill=black] (0,0) circle (0.05);
								\end{scope}
							\end{scope}
						\end{scope}
						\draw[thick] (0,1.7) -- (0,2);
						\begin{scope}[shift={(0,2)}]
							\begin{scope}[scale=0.4]
								\begin{scope}[yscale=0.57735]
									\draw[fill=white!80!blue,postaction={decorate},
									decoration={markings, mark= at position 0.65 with {\arrow{stealth}}}] (0,0) circle (1);
									\draw[dashed, postaction={decorate},
									decoration={markings, mark= at position 0.65 with {\arrow{stealth}}}] (0,0) circle (0.7);
									\draw[dashed, postaction={decorate},
									decoration={markings, mark= at position 0.65 with {\arrow{stealth}}}] (0,0) circle (0.4);
									\draw[fill=black] (0,0) circle (0.05);
								\end{scope}
							\end{scope}
						\end{scope}
						\draw[-stealth, thick] (0,2) -- (0,2.4);
						\node[left] at (0,2.4) {$\scriptstyle h$}; 
						\draw[-stealth, thick] (-1.4,-0.3) -- (-1.4,2);
						\node[left] at (-1.4,2) {$\scriptstyle \vec B$};
					\end{tikzpicture}
				\end{array}
				$};
			\node (C) at (4,-2.5) {$
				\begin{array}{c}
					\begin{tikzpicture}[rotate=-120]
						\begin{scope}[scale=0.7]
							\draw[thick, -stealth] (0,0) -- (0.866025, -0.5);
							\node[left] at (0.866025, -0.5) {$\scriptstyle -{\rm Re}\,\sigma$};
							\draw[thick, -stealth] (0,0) -- (-0.866025, -0.5);
							\node[left] at (-0.866025, -0.5) {$\scriptstyle {\rm Im}\,\sigma$};
						\end{scope}
						\draw[thick,-stealth] (0,0) -- (0,3);
						\node[right] at (0,3) {$\scriptstyle h$};
						\begin{scope}[shift={(0,0.3)}]
							\begin{scope}[scale=0.27]
								\foreach \a/\b in {0/0, 1/0, 1/1, 0/1}
								{
									\draw[fill=\myblue] (0.866025*\a -0.866025*\b , -0.5*\a-0.5*\b) -- (0.866025*\a -0.866025*\b + 0.866025, -0.5*\a-0.5*\b-0.5) -- (0.866025*\a -0.866025*\b, -0.5*\a-0.5*\b-1) -- (0.866025*\a -0.866025*\b -0.866025, -0.5*\a-0.5*\b -0.5) -- cycle;
								}
							\end{scope}
						\end{scope}
						\begin{scope}[shift={(0,0.8)}]
							\begin{scope}[scale=0.27]
								\foreach \a/\b in {0/0, 1/0, 1/1, 0/1, 2/0, 0/2}
								{
									\draw[fill=\myblue] (0.866025*\a -0.866025*\b , -0.5*\a-0.5*\b) -- (0.866025*\a -0.866025*\b + 0.866025, -0.5*\a-0.5*\b-0.5) -- (0.866025*\a -0.866025*\b, -0.5*\a-0.5*\b-1) -- (0.866025*\a -0.866025*\b -0.866025, -0.5*\a-0.5*\b -0.5) -- cycle;
								}
							\end{scope}
						\end{scope}
						\begin{scope}[shift={(0,1.5)}]
							\begin{scope}[scale=0.27]
								\foreach \a/\b in {0/0,0/1,0/2,0/3, 1/0,1/1,1/2, 2/0,2/1,2/2, 3/0,3/1, 4/0,4/1, 5/0}
								{
									\draw[fill=\myblue] (0.866025*\a -0.866025*\b , -0.5*\a-0.5*\b) -- (0.866025*\a -0.866025*\b + 0.866025, -0.5*\a-0.5*\b-0.5) -- (0.866025*\a -0.866025*\b, -0.5*\a-0.5*\b-1) -- (0.866025*\a -0.866025*\b -0.866025, -0.5*\a-0.5*\b -0.5) -- cycle;
								}
							\end{scope}
						\end{scope}
						\begin{scope}[shift={(0,2)}]
							\begin{scope}[scale=0.27]
								\foreach \a/\b in {0/0,0/1,0/2, 1/0,1/1}
								{
									\draw[fill=\myblue] (0.866025*\a -0.866025*\b , -0.5*\a-0.5*\b) -- (0.866025*\a -0.866025*\b + 0.866025, -0.5*\a-0.5*\b-0.5) -- (0.866025*\a -0.866025*\b, -0.5*\a-0.5*\b-1) -- (0.866025*\a -0.866025*\b -0.866025, -0.5*\a-0.5*\b -0.5) -- cycle;
								}
							\end{scope}
						\end{scope}
						\begin{scope}[shift={(0,2.5)}]
							\begin{scope}[scale=0.27]
								\foreach \a/\b in {0/0}
								{
									\draw[fill=\myblue] (0.866025*\a -0.866025*\b , -0.5*\a-0.5*\b) -- (0.866025*\a -0.866025*\b + 0.866025, -0.5*\a-0.5*\b-0.5) -- (0.866025*\a -0.866025*\b, -0.5*\a-0.5*\b-1) -- (0.866025*\a -0.866025*\b -0.866025, -0.5*\a-0.5*\b -0.5) -- cycle;
								}
							\end{scope}
						\end{scope}
					\end{tikzpicture}
				\end{array}
				$};
			\draw[-stealth] (-2.5,0) to[out=180,in=90] node(D) [pos=0.5,above left] {branch} (-4,-0.8);
			\draw[-stealth] (2.5,0) to[out=0,in=90] node(E) [pos=0.5,above right] {branch} (4,-0.8);
			\node[above] at (D.north) {Coulomb};
			\node[above] at (E.north) {Higgs};
		\end{tikzpicture}
	\end{center}
	\caption{An effective molecular picture of localization to the Coulomb and Higgs branches. 
	A diagram in the center is simply an abstract depiction of a molecule inspired by the caffeine molecular structure.}\label{fig:Higgs-Coulomb}
\end{figure}
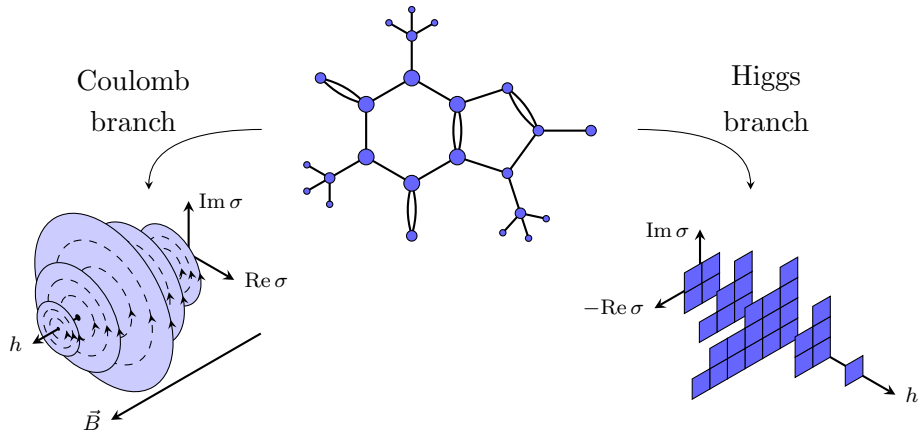

In the case of the Coulomb branch (see discussion in \cite{Galakhov:2018lta}) slow IR degrees of freedom are eigen values of coordinate triplet $({\rm Re}\,\sigma,{\rm Im}\,\sigma,h)$ matrices.
One could deform the Morse height function in such a way that this deformation mimics an external magnetic field directed parallel to the $h$-axis.
So, effectively, on the Coulomb branch in the IR the system is represented by a gas of particles in 3d space parameterized by  $({\rm Re}\,\sigma,{\rm Im}\,\sigma,h)$.
The particles are gathered in the $h$-direction in groups depending on the form of the effective one-loop potential, whereas in the complex $\sigma$-plane the particles occupy Landau levels (see a depiction in fig.~\ref{fig:Higgs-Coulomb}).
Mathematically the IR wave function is an element of corresponding equivariant cohomology -- cohomology of the gauge group classifying space \cite{KS}.
Elements of the cohomology ring can be parameterized by complex field $\sigma$ \cite{Galakhov:2018lta}, where the action of remnant symmetry after symmetry breaking acts on $\sigma$-egenvalues by permutations.
Thus, similarly, to the Laughlin wave function ansatz in the external magnetic field the effective IR wave-functions are given by symmetric polynomials:
\begin{equation}\label{Coul_wave}
	\mathscr{H}_{\rm BPS}^{\rm Coulomb}\cong \prod\lm_{i\in\fQ_0}H^*(BGL(n_i,\IC))\cong\prod\lm_{i\in\fQ_0}\IC\left[{\rm Tr}\,\sigma_i,\, {\rm Tr}\,\sigma_i^2,\,{\rm Tr}\,\sigma_i^3,\ldots\right]\,.
\end{equation}

For quivers associated with toric Calabi-Yau three-folds (our primary interest is aimed for) the structure of the Higgs (mixed) branch is peculiar as well.
Not only fields $\phi_a$ acquire expectation values, rather it is useful to organize those field in quiver path operators \cite{Galakhov:2020vyb, Aspinwall:2009isa} according to a concatenation rule of the quiver path algebra.
Those operators that acquire vevs for a stable BPS state turn out to be labeled solely by the equivariant flavor weight and the R-charge.
Possible quantum numbers of those operators form a 3d crystal appearing in a crystal melting model for DT invariants \cite{mozgovoy2010noncommutative, Ooguri:2009ijd, Aganagic:2010qr, Yamazaki:2010fz}.
We will not go over details of the molten crystal construction.
The reader is encouraged to review an expository summary in \cite{Yamazaki:2010fz, Yamazaki:2022cdg, Galakhov:2021vbo}.
If one considers a brane of a lower dimension wrapping a divisor inside a brane covering the whole CY${}_3$, for example, a $\IC^2$-cycle inside $\IC^3$ \cite{Rapcak:2018nsl, Rapcak:2021hdh, Galakhov:2021xum, Noshita:2021dgj} one could cut out of the 3d crystal a 2d slice of one level points in, say, the R-charge direction.
In this case eigen values of $\sigma$-fields acquire expectation values corresponding to flavor weights of path operators in the complex $\sigma$-plane.
We could picture this regime in a form of a crystal in the complex $\sigma$-plane (see fig.\ref{fig:Higgs-Coulomb}).
Comparing two localization pictures we could treat them as two different phases of effective $\sigma$-particles forming a dilute gas in a magnetic trap on the Coulomb branch, whereas on the Higgs branch those particles ``freeze" in a form of crystal slices.

Despite those two descriptions are rather different one expects that those pictures are dual.
This duality is rather non-trivial and in higher dimensions is related to spectral duality \cite{Gaiotto:2013bwa} and mirror symmetry \cite{Aganagic:2001uw, Dimofte:2019zzj}.
Higgs-Coulomb duality for so called scaling states, or 3d crystal states having distinguishable baryon vevs with the same flavor weight -- hovering over the same position in the $\sigma$-plane and having different R-charge eigen values -- seems less transparent.
In the picture depicted in the right hand side of fig.~\ref{fig:Higgs-Coulomb} this would look like we are trying to put more than one atom in a single crystal cell.
However recent developments \cite{Beaujard:2021fsk} indicate that an attentive counting of R-charge corrections may be helpful to resolve those difficulties. 
For our convenience and simplicity we would not consider those issues in this text assuming in the crystal phase the states are represented by 2d crystals.
As it was shown in \cite{Galakhov:2021xum, Noshita:2021dgj} the crystal dimensionality does not affect the corresponding BPS algebra much, delivering so called ``shifts".

The Higgs-Coulomb duality reveals itself in the most manifest form when one considers various types of partition functions.
The duality implies that the result of the partition function calculation -- a real or a complex number depending on parameters of the system -- is independent of the localization way.
By localizing first to the Coulomb branch one ends up with (see e.g. \cite{Ohta:2015fpe}) an integral over Cartan subalgebra of the gauge group, in other words over $\sigma_i$.
This integration requires a choice of integration contour and is usually referred to as a Jeffrey-Kirwan \cite{jeffrey1995localization} residue prescription choice.
It leads eventually to enumerating integration cycles. Enumeration runs over poles in the $\sigma$-plane these cycles encircle.
Those are positions of $\sigma_i$ as they appear ``frozen'' in the Higgs crystal phase.


\section{BPS algebra from generalized cohomology} \label{s:GCT}


\subsection{Generalized cohomology, cobordism classification and formal group laws}

Naturally generalized (Eilenberg-Steenrod) cohomology theory $E^*$ emerges \cite{may1999concise} when starting with an abstract axiomatic definition of a cohomology theory one relaxes a requirement that an Abelian group $E^k({\rm pt})$ is necessarily $\IZ\, \delta_{k,0}$.
In general \cite{eilenberg1945axiomatic}, $E^*$ is a functor from pairs of topological spaces $(X,A)$ to a graded Abelian group satisfying ordinary cohomology axioms except the dimension axiom:
\begin{enumerate}
	\item \emph{Homotopy invariance.} If $f:\;(X,A)\to (Y,B)$ is a homotopy equivalence then it induces an isomorphism on cohomology:
	\begin{equation}
		f^*:\;\;E^*(Y,B)\longrightarrow E^*(X,A)\,.
	\end{equation}
	\item \emph{Additivity.}  For a disjoint union $(X,A)=\coprod_i(X_i,A_i)$ one has an isomorphism:
	\begin{equation}
		E^n(X,A)\cong\bigoplus\lm_iE^n(X_i,A_i)\,.
	\end{equation}
	\item \emph{Excision.} If $U\hookrightarrow A\hookrightarrow X$ so that closure $\bar U$ is in an interior of $A$ then the inclusion map $\iota:\;(X\setminus U,A\setminus U)\hookrightarrow (X,A)$ induces an isomorphism in cohomology:
	\begin{equation}
		\iota^*:\;\;E^*(X,A)\longrightarrow E^*(X\setminus U,A\setminus U)\,.
	\end{equation}
	\item \emph{Exactness.} Natural inclusions $i:\;A\to X$ and $j:\;(X,\varnothing)\to (X,A)$ induce a long exact sequence:
	\begin{equation}\label{axiom}
		\ldots\; \longrightarrow \;E^n(X,A)\;\overset{j^*}{\longrightarrow} \;E^n(X)\;\overset{i^*}{\longrightarrow}\; E^n(A)\;\longrightarrow E^{n+1}(X,A)\;\longrightarrow\;\ldots\,.
	\end{equation}
\end{enumerate}

A good working model for generalized (co)homology is a (co)bordism ring \cite{Freed_lectures} where continuous maps of simplices are substituted by continuous maps of manifolds graded by their dimension.
Two manifolds $M$ and $M'$ of dimension $d$ with extra topological structure $\fB$ are said to be cobordant if their disjoint union $M\sqcup M'$ is a boundary of a $d+1$-dimensional manifold $W$ and $\fB$ can be smoothly extended to $W$.
The cobordism relation is apparently an equivalence relation, so one can construct a space of all structure $\fB$ cobordism equivalence classes $\Omega_*^{\fB}$.
It is graded by the manifold dimension and has a ring structure induced by disjoint union and Cartesian product operations on the class representatives:
\begin{equation}
	\left[X\sqcup Y\right]=[X]+[Y],\quad \left[X\times Y\right]=[X]\cdot [Y]\,.
\end{equation}

Pontryagin-Thom isomorphism \cite{pontryagin1938classification,thom1954quelques,milnor1974characteristic} identifies $\Omega_*^{\fB}$ with homotopy groups of a Thom space and, further, a Thom spectrum associated to structure $\fB$:
\begin{equation}
	\Omega_*^{\fB}\cong \pi_*\left(M\fB\right)\,.
\end{equation}
A nice practical corollary of this theorem we are planning to implement is a possibility to classify manifolds: to identify a cobordism class of a given manifold using characteristic classes.
In what follows we are planning to consider complex manifolds and their cobordism classes $\Omega_*^U$ with respect to a stably complex structure.
In general, ring $\Omega_*^U$ is rather involved due to multiple torsion elements.
However if one ``erases'' those elements by tensoring with rationals $\IQ$ (considering a ring homomorphism $\Omega_*^U\to \Omega_*^U\otimes \IQ$), or rather we would tensor it with $\IC$ to match the structure field of the most common physical Hilbert spaces, the ring is generated by cobordism classes of projective spaces $[\IC\IP^k]$ \cite{Quillen,quillen1971elementary}:
\begin{equation}
	\Omega_*^U\otimes \IC\cong\IC\left[[\IC\IP^1],[\IC\IP^2],[\IC\IP^3],[\IC\IP^4],\ldots\right]\,,
\end{equation}
In what follows we will extensively use ring $\Omega_*^U\otimes \IC$ as a model for our generalized cohomology theory, therefore to abbreviate notations by label $E^*$ we always imply $\Omega_*^U\otimes \IC$.

We can construct a generic classifying homomorphism called a generalized genus \cite{ochanine2009elliptic, Quillen}:
\begin{equation}
	\xi:\;\;E^*\;\longrightarrow\; \Lambda\,,
\end{equation}
where $\Lambda$ is a universal Lazard's ring that we could identify with $E^*({\rm pt})$.
Having homomorphism $\xi$ a standard way to proceed is to construct a generating function for $E^*$ generators usually called a \emph{logarithm} of $\xi$:
\begin{equation}\label{log}
	\log_{\xi}z:=\sum\lm_{k=0}^{\infty}\frac{\xi\left(\IC\IP^k\right)}{k+1}z^{k+1}\,.
\end{equation}
An inverse function ${\bf e}_{\xi}$ for the logarithm, so that ${\bf e}_{\xi}\left(\log_{\xi}z\right)=\log_{\xi}{\bf e}_{\xi}(z)=z$, we call an exponent of $\xi$ by analogy.
Then for a generic differentiable complex manifold $M$ map $\xi$ is given by the Hirzebruch genus formula \cite{hirzebruch1966topological}:
\begin{equation}\label{hirzebruch}
	\xi(M)=\int\lm_{M}\prod\lm_{j}\frac{u_j}{{\bf e}_{\xi}(u_j)}\,,
\end{equation}
where $u_j$ are Chern roots
for the cotangent bundle of $M$.

An integrand of the Hirzebruch formula \eqref{hirzebruch} is a ring isomorphism\footnote{Apparently this is not an isomorphism from the original ring $\Omega^U_*$ rather a homomorphism since we send all the torsion elements to 0.} due to the Pontryagin-Thom isomorphism:
\begin{equation}\label{GCT_iso}
	\Xi:\;\;E^*(M)\;\longrightarrow \;\Lambda\otimes H^*(M,\IC)\,.
\end{equation}
Indeed, we could calculate the expansion explicitly:
\begin{equation}\label{Xi}
\begin{split}
	\Xi=&1+\frac{c_1 \xi _1}{2}+\frac{1}{12} \left(c_1^2 \left(4 \xi _2-3 \xi _1^2\right)+c_2 \left(9 \xi _1^2-8 \xi _2\right)\right)+\\
	&+\frac{1}{24} \left(6 c_1^3 \left(\xi _1^3-2 \xi _2 \xi _1+\xi _3\right)+c_2 c_1 \left(-21 \xi _1^3+40 \xi _2 \xi _1-18 \xi _3\right)+6 c_3 \left(5 \xi _1^3-8 \xi _2 \xi _1+3 \xi _3\right)\right)+\ldots\,,
\end{split}
\end{equation}
where $\xi_k:=\left[\IC\IP^k\right]$, and $c_k$ is the  $k^{\rm th}$ Chern class.
Substituting explicitly Chern classes for $\IC\IP^k$ we find immediately that:
\begin{equation}
	\Xi\left(\IC\IP^k\right)\Big|_{H^{2k}(\IC\IP^k)}=\xi(\IC\IP^k)\,\omega^k\,,
\end{equation}
where $\omega$ is a generator of $H^{2}(\IC\IP^k;\IZ)$ -- the K\"ahler form in terms of the Fubini-Study metric divided by $2\pi$.

Thom space $MU(1)$ is homotopically equivalent to the classifying space $BU(1)\cong\IC\IP^{\infty}$ \cite{adams1974stable}.
The first Chern class of a complex line bundle $L$ over $X$ is a homotopy class of maps $f:\,X\to BU(1)$.
Combining with homotopy equivalence $BU(1)\cong MU(1)$ one would acquire a class in $E^*$.
Such generalized Chern classes where proposed by Conner and Floyd \cite{adams1974stable, Conner}.
We could define the first Conner-Floyd Chern class as:
\begin{equation}\label{Conner-Floyd}
	c_1^{E}(L)={\bf e}_{\xi}(c_1(L))\in E^*({\rm pt})\otimes H^*(BU(1),\IC)\,,
\end{equation}
indeed for \eqref{Xi} we have:
\begin{equation}
	\Xi=\frac{c_1}{{\bf e}_{\xi}(c_1)}+\sum\lm_{k\geq 2}c_k P_k\,,
\end{equation} 
where $P_k$ are formal series in $c_j$.
Then the cobordism class (the Hirzebruch genus) of a cycle dual to ${\bf e}_{\xi}(c_1)\wedge\alpha_X$ where $\alpha_X$ is dual to a cycle produced by $f:\,X\to BU(1)$ is exactly the ordinary first Chern class.
See also a more detailed discussion in \cite{Atlas}.

Due to a relation for Chern classes for a product bundle $c_1(L\otimes L')=c_1(L)+c_1(L')$ the modified Chern classes satisfy a modified multiplication formula called a formal group law $F_{\xi}$:
\begin{equation}
	c_1^E(L\otimes L')=F_{\xi}(c_1^E(L),c_1^E(L'))\,,
\end{equation}
where, in our terms:
\begin{equation}
	F_{\xi}(u,v)={\bf e}_{\xi}\left(\log_{\xi}u+\log_{\xi}v\right)=u+v-uv\left[\IC\IP^1\right]+uv(u+v)\left(\left[\IC\IP^1\times\IC\IP^1\right]-\left[\IC\IP^2\right]\right)+\ldots\,.
\end{equation}

Since we will not exploit the mechanism of formal group laws to its full generality always using explicit maps ${\bf e}_{\xi}$ and $\log_{\xi}$ instead, the interested reader could become acquainted with a more generic discussion of formal group laws applied to cobordism rings in \cite{Atlas,Lazard,Quillen,MR423332}.


\subsection{Equivariant localization}\label{sec:localization}

A theory of equivaraint cobordisms for finite groups is discussed in \cite{dieck1970bordism}. 

Having a $G$-manifold $M$ it is natural to talk about its $G$-equivariant cohomologies using classifying space $BG$ to ``smear'' fixed points $M^G$ of the $G$-action \cite{Cordes:1994fc, atiyah1984moment}.
In a similar fashion we could proceed to a equivariant generalized cohomology theory $E^*_G(M)$.
Having a contractible universal $G$-bundle $EG\to BG$ over the classifying space \cite{Equivariant_sheves, madsen2016classifying} we could construct a smeared version:
\begin{equation}
	M\times_G EG:=\frac{M\times EG}{G}\,,
\end{equation}
where the $G$-action in local coordinates has the following form $g\cdot(x,u)=(gx,u g^{-1})$.
It would be natural to simply identify $E^*_G(M)$ with $E^*(M\times_G EG)$, however the latter space is infinite dimensional and therefore non-compact.
A nice safety maneuver on this route would be to implement first map \eqref{GCT_iso} to characteristic classes and substitute the latter by respective equivariant versions, so that eventually the equivariant generalized cohomology is a functor from topological complex spaces to rings:
\begin{equation}\label{equiv_funct}
	E_G^*:\;\;M\;\longrightarrow\;E^*({\rm pt})\otimes H_G^*(M;\IC)\,.
\end{equation}

Actually, for our purposes to study eventually quiver moduli spaces a notion of an algebraic variety should be implemented in the construction as well.
A suitable cohomology theory is a Borel-Moore cohomology theory of algebraic cobordisms introduced by Levine and Morel \cite{levine2001cobordisme, levine2007algebraic} (see also a version by Gepner and Snaith \cite{gepner2009motivic}) intertwining notions of generalized cohomology and of an algebraic variety.
We expect that this functor would have similar properties to functor \eqref{equiv_funct}.

A drastic simplification brought in calculations by an adjective ``equivariant" is a version of localization theorem expected from the corresponding theory.
In the language of algebraic varieties localization is naturally expected from such generic objects as derived categories \cite{HL_cat} not mentioning it for algebraic cobordisms.
Following the canonical Atiyah-Bott localization procedure \cite{atiyah1984moment} (see also \cite{guillemin2013supersymmetry}) we would consider a union of small $G$-invariant neighborhoods $U$ for $G$-action fixed points $M^G$ and a long exact sequence delivered to us by one of the cohomology theory axioms \eqref{axiom} for natural inclusion $\iota:\,U\to M$:
\begin{equation}
	\ldots\; \longrightarrow \;E_G^n(M,U)\;\longrightarrow\;E^n_G(M)\;\overset{\iota^*}{\longrightarrow}\; E_G^n(U)\;\longrightarrow E_G^{n+1}(M,U)\;\longrightarrow\;\ldots\,.
\end{equation}
Then we would argue that $E_G^*(M,U)$ are purely torsion modules for $E^*_G$.
In our torsion-free setting map $\iota^*$ is an isomorphism since it has zero kernel and cokernel.

On the other hand we could put forward a more physical argument referring to global properties of differential forms.
This argument represents a localization procedure \`a la Berline-Vergne route \cite{berline1982classes} (see also \cite[ch. 10.10]{guillemin2013supersymmetry}).
Let us restrict ourselves to a simple circle action $G=U(1)$.
This action creates on $M$ Killing vector field $v=v^i\frac{\p}{\p x^i}$.
Then in the Cartan model of equivariant cohomology the differential reads \cite{Cordes:1994fc, witten1982supersymmetry}:
\begin{equation}
	d_G=d+u\,\iota_v\,,
\end{equation}
where $u$ is a degree 2 generator of $H^2_{U(1)}({\rm pt})$ -- symmetric generator of the even Weyl algebra part $S(\fu(1))$ -- a ``curvature" of the universal $EU(1)\to BU(1)$ bundle.
Consider a globally well-defined on $M\setminus U$ form
\begin{equation}
	\psi:=\frac{g_{\mu\nu}v^{\mu}dx^{\nu}}{|\vec v|^2}\,,
\end{equation}
where $g_{\mu\nu}$ is a Riemannian metric on $M$.
This form satisfies two constraints:
\begin{equation}
	\iota_v\psi=1,\quad \CL_v\psi=(v^{\nu}dx^{\mu}-v^{\mu}dx^{\nu})\nabla_{\nu}\frac{v_{\mu}}{|\vec v|^2}=0\,,
\end{equation}
where $\nabla_{\mu}$ is a covariant derivative, and we used metric covariance $\nabla_{\mu}g_{\lambda\rho}=0$ and a Killing constraint $\nabla_{\mu}v_{\nu}+\nabla_{\nu}v_{\mu}=0$.
Using this form we construct another globally defined on $M\setminus U$ form:
\begin{equation}
	\Psi:=\frac{\psi}{d_G\psi}=\frac{\psi}{u+d\psi}=\frac{\psi}{u}\left(1-\frac{d\psi}{u}+\left(\frac{d\psi}{u}\right)^2-\left(\frac{d\psi}{u}\right)^3+\ldots\right)\,.
\end{equation}
Using this new form it is rather simple to show that any $d_G$-closed form $\alpha$ (including equivaraint characteristic classes) is $d_G$-exact:
\begin{equation}
	d_G(\Psi\wedge\alpha)=d_G\Psi\wedge \alpha=\frac{d_G\psi}{d_G\psi}\wedge\alpha=\alpha\,.
\end{equation}
Thus we conclude that any classifying characteristic class on $M\setminus U$ and, therefore, $E_G^*(M)$ is accumulated only in a vicinity of $G$-fixed submanifold $M^G$.

It is natural to define an ``integration'', or an evaluation map, in this cohomology theory as a computation of an equivariant generalized genus, in other words we should calculate an equivariant integral of the Hirzebruch integrand \eqref{hirzebruch}:
\begin{equation}
	\int\Xi\wedge:\;\;E_G^{*}(M)\;\longrightarrow\;E^*_G({\rm pt})\cong E^*({\rm pt})\otimes H^*(BG)\,.
\end{equation}
If the $G$-action is given by a torus $T$-action with weights $u_j$ it is simple to calculate this integral explicitly using the Atiyah-Bott-Berline-Vergne (ABBV) localization formula:
\begin{equation}\label{localization}
	\int \Xi\wedge\alpha=\sum\lm_p\prod\lm_{j}\frac{u_j^{(p)}}{{\bf e}_{\xi}(u_j^{(p)})}\times\frac{\iota^*_p\alpha\left(u_1^{(p)},u_2^{(p)},\ldots\right)}{\prod\lm_j u_j^{(p)}}=\sum\lm_p\frac{\iota^*_p\alpha\left(u_1^{(p)},u_2^{(p)},\ldots\right)}{\prod\lm_j {\bf e}_{\xi}(u_j^{(p)})}\,,
\end{equation}
where the summation runs over fixed points of the $T$-action on $M$ and $u_j^{(p)}$ are $T$-action weights in corresponding point $p$.
In other words we apply the usual ABBV formula with Chern characters of line bundles associated with torus weights substituted  by corresponding Conner-Floyd characters \eqref{Conner-Floyd}.

This ``integration" map coincides with a push-froward map for Grassmann bundles in \cite{MR3805051, HornbostelKiritchenko}.


\subsection{BPS algebra from a mathematical viewpoint}\label{sec:alg_math}

Historically one could distinguish two possible approaches to the construction of BPS algebras from the cohomological data as discussed above for the quiver theories.

The first approach to the BPS algebras returns to the original definition of BPS algebras via a scattering S-matrix \cite{Harvey:1995fq}.
Kontsevich and Soibelman \cite{KS} proposed a construction of an algebra as a cohomological Hall algebra (CoHA) \cite{2006math.....11617S} leading to a multiplication shuffle algebra in rings of wave functions \eqref{Coul_wave}.
Naturally one might notice that there is a natural embedding of complexified gauge groups:
\begin{equation}\label{emb}
	\prod\lm_{i\in \fQ_0} GL(n_i,\IC)\times\prod\lm_{i\in \fQ_0} GL(m_i,\IC)\;\longrightarrow\;\prod\lm_{i\in \fQ_0} GL(n_i+m_i,\IC)\,,
\end{equation}
that induces a natural embedding of rings \eqref{Coul_wave}.
As a result the multiplication in the BPS algebra (CoHA) maps a pair of wave functions on the Coulomb branch for quiver reps with dimension vectors $\{n_i\}_{i\in\fQ_0}$ and $\{m_i\}_{i\in\fQ_0}$ to a wave function for a sum vector $\{n_i+m_i\}_{i\in\fQ_0}$.
The calculation of effective IR wave functions has a natural generalization in the setting of generalized cohomology \cite{MR3805051}.
An extra form factor appearing due to IR renormalization of off-diagonal degrees of freedom in embedding \eqref{emb} can be re-organized in a form of Euler classes \cite{Galakhov:2018lta}.
Switching from ordinary cohomology theory to a generalized cohomology theory leads to a substitution of Euler classes by generalized Euler classes ${\bf e}_{\xi}$.
It is natural to ask if the resulting modified product is still a valid product for the ring of symmetric functions \eqref{Coul_wave}, in particular, if symmetric polynomials are mapped to symmetric polynomials.
A proof of the positive answer to this question could be found in \cite{HornbostelKiritchenko}.
We will discuss a physical construction and an explicit formula for multiplication in this CoHA associated with a generalized cohomology theory in sec.~\ref{CoHA}.

Now we turn to an earlier idea of Nakajima \cite{10.1215/S0012-7094-94-07613-8,nakajima1999lectures} based on an application of Hecke modifications to instanton moduli spaces adding or subtracting instantons (see also a construction for vortex moduli spaces \cite{Bullimore:2016hdc, Braverman:2016wma}).
The instanton moduli space as well as a Hilbert scheme on $\IC^2$ have an ADHM description in terms of a quiver representation.
The Hecke modification increases/decreases the instanton numbers, in the quiver language these numbers are quiver dimensions.
The resulting operators are raising or lowering operators corresponding to each gauge node, they correspond to embeddings:
\begin{equation}\label{embedding}
	\prod\lm_{i\in\fQ_0}GL(n_i,\IC)\;\longrightarrow\;\prod\lm_{i\in\fQ_0}GL(n_i+\delta_{i,k},\IC)\,,\quad k\in\fQ_0\,.
\end{equation}
Surely, one may consider a more complicated embedding when dimension vectors differ in a more complicated way than just a unit vector.
However to construct those higher operators one imposes more complicated constraints, in particular, the way one complexified gauge group is embedded into the other.
An explicit representation for these operators could be constructed naturally in the language of the Higgs branch as we described it -- in terms of crystals.

Before writing down expressions for generators let us introduce some notions first.

Consider two quiver representations $\CR$ and $\tilde \CR$.
We will denote used quantities such as dimensions $n_{i\in\fQ_0}$, vector spaces $V_{i\in \fQ_0}$ associated to quiver nodes and morphisms $\phi_{(a:i\to j)\in \fQ_1}\in {\rm Hom}(V_i,V_j)$ corresponding to $\CR$ and $\tilde \CR$  accordingly by letters with a tilde or without.
A homomorphism of representations $\CR\sim\tilde \CR$ is a set of maps $\{\tau_i\}_{i\in\fQ_0}$ such that the following diagram commutes for all arrows $a\in \fQ_1$:
\begin{equation}
	\begin{array}{c}
		\begin{tikzpicture}
			\node(A) at (0,0) {$V_i$};
			\node(B) at (3,0) {$V_j$};
			\node(C) at (0,-1.5) {$\tilde V_i$};
			\node(D) at (3,-1.5) {$\tilde V_j$};
			\path (A) edge[->] node[above] {$\scriptstyle \phi_{a:i\to j}$} (B) (C) edge[->] node[above] {$\scriptstyle \tilde\phi_{a:i\to j}$} (D) (A) edge[->] node[left] {$\scriptstyle \tau_i$} (C) (B) edge[->] node[right] {$\scriptstyle \tau_j$} (D);
		\end{tikzpicture}
	\end{array}\,.
\end{equation}

Now assume that $\CR$ and $\tilde \CR$ correspond to $v=\{n_i+\delta_{i,k}\}_{i\in\fQ_0}$ and $\tilde v=\{n_i\}_{i\in\fQ_0}$ respectively.
We define an incidence locus $\CI$ as:
\begin{equation}
	\CI:=\left\{(\CR,\tilde \CR)\;\in\; \left({\rm Rep}\,\fQ_0\right)_{v}\times \left({\rm Rep}\,\fQ_0\right)_{\tilde v}\big| \CR\sim\tilde\CR\right\}\,.
\end{equation}
Fixed points on $\CI$ are labeled by pairs of fixed points $\left(\Kappa_1,\Kappa_2\right)$ such that crystal $\Kappa_2$ is embedded in crystal $\Kappa_1$.
Since the dimension vector has a shift 1 at position $k$ the difference between $\Kappa_1$ and $\Kappa_2$ is in a single color $k\in \fQ_0$ atom position somewhere at a vacant place.
We would like to denote such a relation between crystals in the following way:
\begin{equation}
	\Kappa_1=\Kappa_2+\sqbox{$k$},\quad \Kappa_2=\Kappa_1-\sqbox{$k$}\,.
\end{equation}
Let us denote tangent spaces in fixed points as $T\CR_{\Kappa}$ and $T\CI_{\Kappa,\Kappa+\Box}$ respectively.

A Hecke modification algebra for quiver varieties is constructed similarly to \cite{nakajima1999lectures} via a Fourier-Mukai transform for varieties $\left({\rm Rep}\,\fQ_0\right)_{v}$ and $\left({\rm Rep}\,\fQ_0\right)_{\tilde v}$ with a kernel supported at $\CI$ and descended to cohomologies.
To pass to generalized cohomologies we simply substitute characteristic classes by Conner-Floyd classes and integration by the generalized genus.
The algebraic action localizes to fixed points.
We define raising/lowering operator matrix coefficients using the localization formula \eqref{localization}:
\begin{equation}\label{matrix}
	\begin{split}
	&\left[\Kappa\to\Kappa+\Box\right]=\frac{{\bf e}_{\xi}\left(T\CR_{\Kappa}\right)}{{\bf e}_{\xi}\left(T\CI_{\Kappa,\Kappa+\Box}\right)}\,,\\
	&\left[\Kappa\to\Kappa-\Box\right]=\frac{{\bf e}_{\xi}\left(T\CR_{\Kappa}\right)}{{\bf e}_{\xi}\left(T\CI_{\Kappa-\Box,\Kappa}\right)}\,.
	\end{split}
\end{equation}

The approaches of CoHA and Hecke modifications are not unrelated.
The basic element entering relations between matrix elements of the BPS algebra of Hecke modifications is a bond factor $\phi_{i,j}$ (see \eqref{Prm}).
It factorizes in a ratio of a pair of form factors $\eta_{i,j}$ entering the explicit CoHA multiplication formula \eqref{CoHA_eq}.
To pass from the algebra of Hecke modifications to the CoHA one might follow the construction of a shuffle algebra form the Borel positive part of the Hecke modification algebra \cite{FT, Rapcak:2018nsl}.


\section{BPS algebra from instanton/Berry phase counting} \label{s:phys_BPS}


\subsection{BPS Hilbert space as a generalized cohomology theory}

The seminal paper \cite{witten1982supersymmetry} has built a firm bridge connecting vast areas of physics and geometry.
In practice, relations like \eqref{H_BPS} bind physical information about the ground state of a quantum mechanical system to geometric information from the target space of this theory.
By varying various details about the quantum theory like adding extra symmetries one could acquire different cohomology theories.
For instance, in the very \cite{witten1982supersymmetry} there are two cohomology theories described: deRahm cohmology, and the Cartan model for equivariant cohomology.
It is only natural that this topic has acquired various deformations and generalizations upon the present.
However the question if any generalized cohomology theory in the Eilenberg-Steenrod formulation has a clean physical model description is yet open.

In this text we would chase quite more modest ambitions and try to answer the following question:
if and why BPS spectra of quiver gauge theories (or their subsectors) could be described by a generalized cohomology theory.
We base our pursuit on observations of \cite{Galakhov:2021vbo} for BPS spectra of the 4d $\CN=1$ quiver gauge theory compactifications.
Let us separate the temporal direction from the 4d Minkowski space-time and compactify the remaining spacial slice to a resulting space $\Sigma$.
In the cases $\Sigma=$point/circle/torus the resulting BPS  algebra (flux-less subsector of it) is a rational/trigonometric/elliptic version of the quiver Yangian algebra.
To restore the role of the cohomology theory as a description of the BPS Hilbert space let us note that the standard Schr\"odinger picture of a QFT represents field configurations as evolving with time points in the space of maps:
\begin{equation}\label{maps}
	{\rm Map}\left(\Sigma\longrightarrow {\rm Rep}\,\fQ_0\right)\,,
\end{equation}
and the very QFT is considered as a quantum mechanics on space \eqref{maps}.
Eventually it is natural to identify the BPS Hilbert space with cohomologies of \eqref{maps}.
In practice, we are able to identify only a subsector of the BPS Hilbert space with corresponding cohomologies in the case ${\rm dim}\,\Sigma\geq 2$ due to possible non-trivial fluxes and anomalies, see discussions in \cite[sec.~4]{Galakhov:2021vbo} and in \cite{Alvarez-Gaume:1986rcs, Losev:1996up, Closset:2019ucb, Dedushenko:2022pem}.

The supercharge we localize with respect to is similar to \eqref{supercharge} with a significant modification in the vector field part.
Now it includes a diffeomorphism action in addition to gauge and flavor ones:
\begin{equation}\label{vector}
	\bar V=\int\lm_{\Sigma} w\sum\lm_{(a:i\to j)\in \fQ_1}\left(\p_{\bar z}\phi_a^{\dagger}+\phi_a^{\dagger}\sigma_j-\sigma_i\phi_a^{\dagger}-\mu_a\phi_a^{\dagger}\right)\frac{\delta}{\delta\phi_a^{\dagger}}\,,
\end{equation}
where $z$ is a coordinate along the circle in the case $\Sigma=S^1$, and a complex coordinate along $\Sigma$ if $\Sigma$ is a Riemann surface, and $w$ is a volume form.
In the case $\Sigma=\{{\rm pt}\}$ the cohomology theory simply returns to simple cohomology defined by \eqref{supercharge}.
One could put forward arguments that in the three cases $\Sigma=$point/circle/torus the cohomological theories are ordinary cohomology, K-theory and elliptic cohomology respectively.
However in the latter case the major argument is that a function naturally identified with ${\bf e}_{\xi}(\sigma)$ is a Jacobi theta-function.
Moreover the situation becomes even more cumbersome if higher genus Riemann surface is considered for the role of $\Sigma$.

Here we would like to present arguments that for abstractly defined $\Sigma$ (as an example one could bear in mind a family of Riemann surfaces for different genera and moduli as $\Sigma$) the flux-less BPS Hilbert space subsector is described by a cohomology theory for some choice of morphism $\xi$.
The structure of this BPS Hilbert space as well as the structure of the BPS algebra could be described in universal terms independent of $\xi$, in this way we conclude that our BPS Hilbert space is captured by a generalized cohomology theory.

Before turning to explicit relations let us argue why such a conclusion might have been achieved a priori.
The reasoning is two-fold.
The first tremendous simplification takes place when we attach the structure of a vector space to the spectrum of our cohomological (cobordism) theory -- solely an Abelian group from the very beginning.
It seems to be a necessary step since we identify this spectrum with the BPS Hilbert space having an explicit structure of a $\IC$-linear space.
Yet this morphism of Abelian groups maps all the torsion elements to zero.
And the remaining information about the variety in question is captured by characteristic classes.
The second point of the reasoning scheme is to apply the peculiarity of supersymmetry and of the equivariant action induced by the gauge and flavor groups.
This action induces localization so that valuable information about a variety whose cohomology is in question shrinks to a description of a fixed point neighborhood as we discussed in sec.~\ref{sec:localization}.

Summarizing our arguments we could conclude that the BPS Hilbert space is isomorphic to a generalized equivariant cohomology theory as an Abelian group:
\begin{equation}
	\mathscr{H}_{\rm BPS}^*\cong E_G^*\left({\rm Rep}\, \fQ_0\right)\,.
\end{equation}
And the details about what properties of $E^*$ emerge in this model, i.e. what formal group law will take place,  are captured by the field space \eqref{maps}.

To follow this route let us remind that the basic building block of an equivariant cohomology theory is the classifying space $BU(1)\cong MU(1)\cong \IC\IP^{\infty}$.
An effective IR theory of the following Kronecker quiver:
\begin{equation}
	\begin{array}{c}
		\begin{tikzpicture}
			\draw[postaction={decorate},
			decoration={markings, mark= at position 0.65 with {\arrow{stealth}}}] (0,0) to[out=30, in=150] (2,0);
			\draw[postaction={decorate},
			decoration={markings, mark= at position 0.65 with {\arrow{stealth}}}] (0,0) to[out=15, in=165] (2,0);
			\draw[dashed, postaction={decorate},
			decoration={markings, mark= at position 0.65 with {\arrow{stealth}}}] (0,0) to[out=-15, in=195] (2,0);
			\draw[postaction={decorate},
			decoration={markings, mark= at position 0.65 with {\arrow{stealth}}}] (0,0) to[out=-30, in=210] (2,0);
			\draw[fill=\myblue] (2,0) circle (0.1);
			\draw[fill=\myblue] (-0.08,-0.08) -- (-0.08,0.08) -- (0.08,0.08) -- (0.08,-0.08) -- cycle;
		\end{tikzpicture}
	\end{array},\quad \kappa\mbox{ arrows}
\end{equation}
is described by a particle on $\IC\IP^\kappa$.
Eventually one will arrive to the classifying space $BU(1)$ in the limit $\kappa\to\infty$.
Cohomology theory of this space is a ring generated by the first Chern class given by K\"ahler form $\omega$.
Higgs-Coulomb duality \cite{Denef:2002ru} (see also sec.~\ref{sec:HC_du}) identifies rings $\IC[\omega]$ and $\IC[\sigma]$ as effective BPS Hilbert space descriptions on the Higgs and the Coulomb branches respectively.

Assuming that field $\sigma$ in vector field $V$ is external and constant we see that the resulting theory is free.
We could expand field $\phi$ in Fourier momenta $p$ on $\Sigma$.
Effectively $p$ simply shifts $\sigma\to \sigma+p$ due to a diffeomorphism term in \eqref{vector}.
The Euler class of such a theory describes a ground state wave function and is a product of elementary wave functions represented by Euler classes for each $p$:
\begin{equation}\label{regularization}
	{\rm det}\,\iota_{\bar V}=\prod\lm_p\left(\sigma+p\right)\xrightarrow{\hspace*{0.8cm} \zeta{\rm -regularization} \hspace*{0.8cm}}{\bf e}_{\xi}(\sigma)
\end{equation}
for some $\xi$ defined by $\Sigma$.
For $\Sigma$ of higher genera $g>1$ it is given by corresponding Riemann theta-functions \cite{Alvarez-Gaume:1986rcs}.
The reader should be warned that these naive manipulations should be accompanied by various specific details.
For a surface of genus $g>1$ flavor Wilson lines saturating a constant value of $\sigma$ belong to the Jacobian ${\rm Jac}(\Sigma)$, that has a complex dimension $g$.
To restrict $\sigma$ to take values in a complex plane one could confine a consideration of all flavor charges to a plane in ${\rm Jac}(\Sigma)$.
Another issue revealed by the zeta-regularization is an anomaly that breaks either flavor invariance of ${\rm det}\,\iota_{\bar V}$ or its holomorphic behavior lifting it to a section of a determinant bundle \cite{Alvarez-Gaume:1986rcs}.
Eventually for simplicity and to assume $\sigma$ being the constant field, or more precisely belonging to Abelian ${\rm Jac}(\Sigma)$, one has to exclude topologically non-trivial field configurations from the discussion.
This could be achieved by restricting the consideration to a flux-less subsector of the BPS Hilbert space, in other words this subsector does not include possible BPS vortices (see a discussion in \cite[sec.~4]{Galakhov:2021vbo}). 

In this proposal we have intentionally omitted an explicit discussion of Riemann surfaces $\Sigma$ and possible resulting formal group logarithms $\log_{\xi}z$.
By leaving this consideration blank we would like to put forward arguments in this section that work for any $\Sigma$ and allow one to construct a BPS algebra in abstract terms.
As we mentioned before the homomorphism to the cobordism ring over $\IC$ and localization make the consideration of K-theory, elliptic cohomology and generalized cohomology in this model uniform by restricting the most essential information about the model to fixed points of the target space and a genuine function ${\rm det}\,\iota_V$.
Also we should mention that $\left({\rm det}\,\iota_V\right)^{\pm 1}$ is an uncompensated contribution of a one-loop determinant in a holomorphic partition function \cite{Fujimori:2015zaa, Beem:2012mb} giving rise to the elliptic genera.


\subsection{Cohomological Hall algebra}\label{CoHA}

A physical construction of the BPS cohomological Hall algebra (CoHA) \cite{Galakhov:2018lta} is performed in the setting of the Coulomb branch localization.
The basic idea is to translate the mathematical construction of a product of two BPS wave functions $\Psi_1$ and $\Psi_2$ -- elements of ring \eqref{Coul_wave} -- into a picture of two compact molecules separated in the physical space, so that dynamics splits as well into effective dynamics inside each separate molecule, and a mutual interaction between molecules is taken into account as loop corrections.
One is able to achieve such a separation between the centers of mass for these two molecules in the $h$-direction by putting both in a background of another simple heavy core.
As a result the long range interactions between the molecules could be integrated at one loop and lead to a modification of the IR wave function in a form of a form factor we call $\eta$.

We could repeat this procedure for $\Sigma$ by performing the renormalization group integration for each mode individually.
However before presenting the resulting product formula let us first remind the structure of the effective wave function on the Coulomb branch in this framework.
As we mentioned it is an element of ring \eqref{Coul_wave}.
Let us denote eigen values of $\sigma_i$, $i\in\fQ_0$ as $\sigma_i^{\alpha}$, where $\alpha=1,\ldots,n_i$, so that $n_i$ are quiver dimensions.
Therefore a wave function for a stable BPS molecule localized around some $h_1$ is a function:
\begin{equation}
	\Psi_1\left(\{\sigma_1^{\alpha_1}\}_{\alpha_1=1}^{n_1},\{\sigma_2^{\alpha_2}\}_{\alpha_2=1}^{n_2},\ldots\right)\,,
\end{equation}
symmetric in each group of variables $\{\sigma_i^{\alpha_i}\}_{\alpha_i=1}^{n_i}$.
If another molecule corresponding to dimension vector $\{m_i\}_{i\in\fQ_0}$ is placed at $h_2$ (see fig.~\ref{fig:Higgs-Coulomb}) this configuration could be considered as a subsystem of a system with dimension vector $\{n_i+m_i\}_{i\in\fQ_0}$ according to embedding \eqref{emb}.
However the resulting wave function would acquire IR contributions from off-diagonal elements in this embedding.

Let us first consider an arrow $a:i\to j$ connecting nodes $i$ and $j$ assuming $i\neq j$.
Field $\phi_a$ has a component $(\beta,\alpha)$ where $\alpha$ and $\beta$ are parameterizing basis vector indices of $\IC^{n_i}$ and $\IC^{m_j}$.
Quantum field $\left(\phi_a\right)_{\beta\alpha}$ is a function of the coordinate on $\Sigma$ and could be expanded into Fourier modes $\left(\phi_a\right)_{\beta\alpha}(p)$ as a function of corresponding momenta $p$.
Eventually, the equivariant weight of $\left(\phi_a\right)_{\beta\alpha}(p)$ is given by the action of the gauge group, the flavor group and the diffeomorphism part of vector field \eqref{vector}.
According to \eqref{I_3} this contribution reads:
\begin{equation}
	\left(p+\sigma_j^{\beta}-\sigma_i^{\alpha}-\mu_a\right)^{\Theta(h_2-h_1)}\,.
\end{equation}
Let us assign the ordering of the molecules on the $h$-line to the ordering of the corresponding wave functions in the product.
In other words in the product $\Psi_1\cdot\Psi_2$ we assume that for molecules $h_2>h_1$.

In a similar way we expand in modes and take into account contributions of fields in the case of coincident quiver nodes $i=j$, however in this case one acquires an additional contribution of off-diagonal elements of the gauge multiplet due to embedding \eqref{emb} becoming Goldstone particles.
It contributes to the denominator of the IR wave function \cite[sec.~3.2.1]{Galakhov:2018lta}.

Summarizing all the contributions and taking into account that the remnant unbroken Weyl group of the gauge group acts on $\sigma_i^{\alpha}$ by permutations we arrive to the following multiplication formula for the wave functions:
\begin{equation}\label{CoHA_eq}
	\Psi_1\cdot\Psi_2=\mathop{\rm Sym}\lm_{\{\sigma\},\{\sigma'\}}\Psi_1\left(\{\sigma\}\right)\Psi_2\left(\{\sigma'\}\right)\prod\lm_{i,j\in \fQ_0}\prod\lm_{\alpha=1}^{n_i}\prod\lm_{\beta=1}^{m_j}\eta_{i,j}\left(\sigma_j^{\beta}-\sigma_i^{\alpha}\right)\,,
\end{equation}
where an elementary form-factor $\eta$ is given by the following expression:
\begin{equation}\label{eta}
	\eta_{i,j}(z):=\frac{\prod\lm_{a:i\to j}{\bf e}_{\xi}(z-\mu_{a})}{{\bf e}_{\xi}(z)^{\delta_{i,j}}}\,.
\end{equation}
The numerator in this formula is a contribution of chiral fields $(\phi_a)_{\alpha\beta}$, whereas the denominator is a respective contribution of the gauge field off-diagonal elements $(\sigma_i)_{\alpha\neq \beta}$.

This is a straightforward generalization of Kontsevich-Soibelman product formula \cite{KS} and coincides with the generalization proposed in \cite{MR3805051,HornbostelKiritchenko} (to match relations explicitly one should perform a ring automorphism $z\mapsto\log_{\xi} z$ on $\sigma$-variables).


\subsection{Berry connection from an interface}

In this paper we would like to deviate a little bit from the route of \cite{Galakhov:2021vbo} to construct the BPS algebra and try to mention its dynamical origin.
The BPS algebra originates from scattering properties of BPS states \cite{Harvey:1995fq}.
It might happen that it is hard to calculate S-matrices even for BPS states non-perturbatively in some theory, for instance, the issue might be that boosted BPS states do not preserve required supersymmetries anymore, and the S-matrix as a function of Mandelstam variables can not be calculated via localization.
In this case one might hope to circumvent this difficulty by applying the following trick -- to consider instead ``adiabatic" scattering.
By following paths on the moduli space intersecting marginal stability walls one would observe that some BPS particles just decay/recombine naturally, so that their wave function reveal specific behavior \cite{Galakhov:2018lta}.

A nice appropriate tool to include such a type of moduli ``dynamics" into the theory in a natural way is to consider an interface defect.
Interfaces \cite{Gaiotto:2015aoa,Galakhov:2020upa,Dedushenko:2021mds,Bullimore:2021rnr,Brunner:2008fa} introduce a variation of the moduli along some spacial directions and might preserve some supersymmetry subgroup.
After applying the Wick rotation and sending the time direction along the interface one would naturally arrive to a time-dependent (moduli-dependent) Hamiltonian $H(t)$ and a supercharge $Q(t)$.
Thus it is natural to consider the time-dependent BPS Hilbert space and a Berry connection induced on it \cite{Dedushenko:2022pem}.

The time/moduli variation moves the BPS states confined to fixed points on the target space adiabatically, so that if one calculates a BPS wave function $\Psi(\lambda)$ as a function of some constant modulus $\lambda$ then $\Psi(\lambda(t))$ is a good approximation for a time-dependent wave function.
The adiabatic picture breaks down at critical values $\lambda_*$ when the wave functions of states overlap.
In supersymmetric quantum mechanics (or SQFT as we treat it in terms of SQM on space \eqref{maps}) the ground state overlap is controlled by instantons \cite{witten1982supersymmetry}:
\begin{equation}\label{instanton}
	\frac{\p x^I}{\p\tau}=+g^{IJ}\p_J U,\quad \lim\lm_{\tau\to -\infty}x=x_{*\alpha},\;\;\lim\lm_{\tau\to +\infty}x=x_{*\beta}\,,
\end{equation}
where $x^I$ are coordinates on the SQM target-space, $g_{IJ}$ is the metric, and  $x_{*\alpha,\beta}$ are classical vacuum values.
Let us denote the space of one-instantons solving \eqref{instanton} as $\CI_{\alpha,\beta}$.
The supercharge matrix element for the BPS states associated with vacua $\alpha$ and $\beta$ acquires a non-perturbative correction:
\begin{equation}
	\langle \Psi_{\beta}|Q|\Psi_{\alpha}\rangle=\sum\lm_{\gamma\in \CI_{\alpha,\beta}}{\rm det }\,L(\gamma)\, e^{U_{*\beta}-U_{*\alpha}}\,,
\end{equation}
where ${\rm det }\,L(\gamma)$ is a determinant of a Dirac operator $L$ in instanton background $\gamma$ \cite{Gaiotto:2015aoa}, it takes values $\pm 1$.
After the inverse Wick rotation instantons become spatially oriented solitons and contribute to Stockes coefficients of partition function asymptotic jumps in a theory on a manifold with boundary.

In this paper we would like to consider a one instanton contribution in flux-less sectors of quiver gauge theories on $\Sigma\times \IR$.
To simplify the calculation process we would like to implement the following trick by introducing an effective Berry connection along the instanton trajectory.
The Wilsonian renormalization of the BPS wave function is much simpler due to localization, it might turn out to be one-loop exact (see a detailed discussion in \cite[sec.~2.3]{Galakhov:2020vyb}).
Let us consider a single instanton trajectory saturating a particle tunneling process between two wells -- minima of potential $|\nabla U|^2$.
We could modify homotopically height function $U$ in such a way that $|\nabla U|^2$ forms a canyon around the wells and the instanton trajectory:
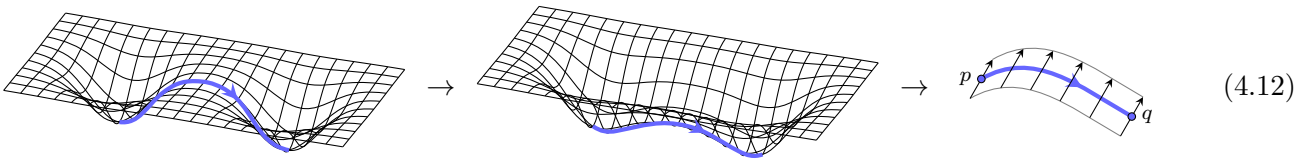
\begin{equation}
	\begin{array}{c}
		\begin{tikzpicture}[scale=1.2]
			\tikzset{myline1/.style = {thin}}
			\tikzset{myline2/.style = {ultra thick, \myblue, postaction={decorate},
					decoration={markings, mark= at position 0.65 with {\arrow{stealth}}}}}
			\input{Misc/graph1.tex}
		\end{tikzpicture}
	\end{array}\to \begin{array}{c}
		\begin{tikzpicture}[scale=1.2]
			\tikzset{myline1/.style = {thin}}
			\tikzset{myline2/.style = {ultra thick, \myblue, postaction={decorate},
					decoration={markings, mark= at position 0.65 with {\arrow{stealth}}}}}
			\input{Misc/graph2.tex}
		\end{tikzpicture}
	\end{array}\to\begin{array}{c}
		\begin{tikzpicture}
			\draw[ultra thick, \myblue, postaction={decorate},
			decoration={markings, mark= at position 0.65 with {\arrow{stealth}}}] (-2,0.5) to[out=30,in=150] node(A1)[pos=0]{} node(A2)[pos=0.2]{} node(A3)[pos=0.4]{} node(A4)[pos=0.6]{} node(A5)[pos=0.8]{} node(A6)[pos=1]{} (0,0);
			\begin{scope}[shift={(0.15, 0.259808)}]
				\draw[thin,gray] (-2,0.5) to[out=30,in=150] (0,0);
			\end{scope}
			\begin{scope}[shift={(-0.15, -0.259808)}]
				\draw[thin,gray] (-2,0.5) to[out=30,in=150] (0,0);
			\end{scope}
			\begin{scope}[shift={(A1.center)}]
				\draw[-stealth] (-0.15, -0.259808) -- (0.15, 0.259808);
			\end{scope}
			\begin{scope}[shift={(A2.center)}]
				\draw[-stealth] (-0.15, -0.259808) -- (0.15, 0.259808);
			\end{scope}
			\begin{scope}[shift={(A3.center)}]
				\draw[-stealth] (-0.15, -0.259808) -- (0.15, 0.259808);
			\end{scope}
			\begin{scope}[shift={(A4.center)}]
				\draw[-stealth] (-0.15, -0.259808) -- (0.15, 0.259808);
			\end{scope}
			\begin{scope}[shift={(A5.center)}]
				\draw[-stealth] (-0.15, -0.259808) -- (0.15, 0.259808);
			\end{scope}
			\begin{scope}[shift={(A6.center)}]
				\draw[-stealth] (-0.15, -0.259808) -- (0.15, 0.259808);
			\end{scope}
			\draw[fill=\myblue] (-2,0.5) circle (0.05) (0,0) circle (0.05);
			\node[left] at (-2,0.5) {$\scriptstyle p$};
			\node[right] at (0,0) {$\scriptstyle q$};
		\end{tikzpicture}
	\end{array}
\end{equation}
This modification does not spoil both the instanton trajectory and the vacua, however it re-scales effective frequencies for the degrees of freedom perpendicular to the trajectory.
We could treat those d.o.f. as ``fast'' variables, whereas d.o.f. parallel to the instanton trajectory are ``slow'' variables, and apply Wilsonian renormalization to this picture.
This procedure induces an effective Berry connection:
\begin{equation}
	B_{\rm IR}=\langle \Psi_{\perp}|d|\Psi_{\perp}\rangle\,,
\end{equation}
where $\Psi_{\perp}$ is the first order calculated BPS wave function, and averaging in the correlator goes over perpendicular d.o.f. only.
Formally, we could say that the Berry connection plays the role of parallel transport for the tangent orientation of the target space along the instanton trajectory. 

Witten's instanton counting in the Morse theory gets corrected by the effective Berry connection on the tangent space for the effective IR wave function $\psi_{\parallel}$:
\begin{equation}
	Q\cdot\psi_{\parallel,p}=\sum\lm_{\gamma\in \CI_{pq}}\langle \psi_{\parallel,q}|{\rm Texp}\,{\int\lm_{\gamma} B_{\rm IR}}|\psi_{\parallel,p}\rangle\cdot\psi_{\parallel,q}\,.
\end{equation}
This trick allows one to determine the sign value of ${\rm det }\,L(\gamma)$ by solely geometric means, see the Morse differential sign rule in \cite[app.F]{Gaiotto:2015aoa} and \cite{witten1982supersymmetry}.
Similarly, we could have calculated perturbatively the Gelfand-Yaglom-van Vleck-Pauli determinant formula \cite{van1928correspondence} as the first order correction in the WKB expansion, again in the case of SQM this first order calculation is exact.


\subsection{Classical R-matrix from an instanton}

Following our plan we would like to calculate one instanton amplitude in our set of models using the trick with the effective Berry connection.
However counting instantons in the quiver SQM or SQFT is not an easy task.

Following \cite{Galakhov:2022uyu} it would be rather spectacular to calculate a one-instanton saturating a migration of an ``atom" between quiver BPS crystals located in the same weight space $(\sigma, h)$.
In this case two crystals located at two values of $h$ represent two vectors of a prospective BPS algebra representations, and embedding them in the same weight space is a physical model for the tensor product.
As it was argued in \cite{Galakhov:2022uyu} one is able to impose different co-products and R-matrices for such a tensor product related by a basis transformation.
The basis transformation is given by a sum over instantons carrying atoms from one crystal to the other, and the first order term in this expansion -- one-instanton amplitude $\mathscr{A}$ -- has a particular form of a classical R-matrix containing the action of raising and lowering operators in the would-be BPS algebra (for a recent discussion of quiver Yangian R-matrix factorization properties see e.g. \cite{Prochazka:2023zdb, Kolyaskin:2022tqi, Bao:2022fpk, Chistyakova:2021yyd, Litvinov:2020zeq} and references therein).
In other words we would be able to divide the instanton amplitude in characteristic pieces corresponding to matrix elements of BPS algebra operators and restore the BPS algebra operators themselves.

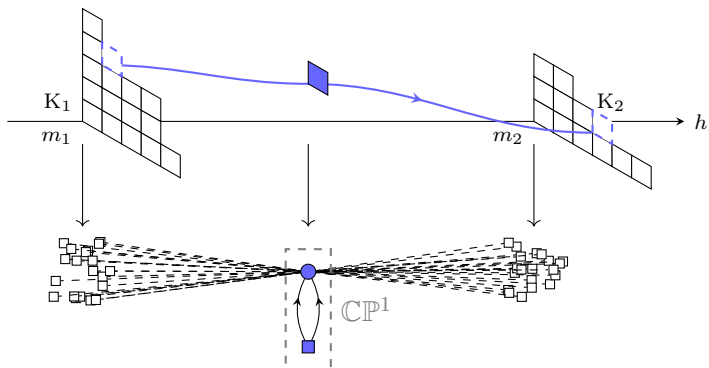
\begin{figure}[h!]
	\begin{center}
		\begin{tikzpicture}
			\draw [-stealth] (-4,0) -- (5,0);
			\node[right] at (5,0) {$\scriptstyle h$};
			\begin{scope}[shift={(-3,0)}]
				\begin{scope}[scale=0.3]
					\foreach \x/\y in {1/3}
					{
						\node(A) at (0.866025*\x, -0.5*\x + \y) {};
					}
				\end{scope}
			\end{scope}
			\begin{scope}[shift={(3,0)}]
				\begin{scope}[scale=0.3]
					\foreach \x/\y in {3/1}
					{
						\node(B) at (0.866025*\x, -0.5*\x + \y) {};
					}
				\end{scope}
			\end{scope}
			\begin{scope}[shift={(3,0)}]
				\begin{scope}[scale=0.3]
					\foreach \x/\y in {0/0, 0/1, 0/2, 1/0, 1/1, 1/2, 2/0, 2/1, 3/0, 4/0 ,5/0}
					{
						\draw[fill=white] (0.866025*\x, -0.5*\x + \y) -- (0.866025*\x, -0.5*\x + \y + 1) -- (0.866025*\x + 0.866025, -0.5*\x + \y + 0.5) -- (0.866025*\x + 0.866025, -0.5*\x + \y -0.5) -- cycle;
					}
					\foreach \x/\y in {3/1}
					{
						\draw[thick, dashed, \myblue, fill=white] (0.866025*\x, -0.5*\x + \y) -- (0.866025*\x, -0.5*\x + \y + 1) -- (0.866025*\x + 0.866025, -0.5*\x + \y + 0.5) -- (0.866025*\x + 0.866025, -0.5*\x + \y -0.5) -- cycle;
					}
				\end{scope}
			\end{scope}
			\draw[\myblue, thick, postaction={decorate}, decoration={markings, mark= at position 0.65 with {\arrow{stealth}}}] (A.center) to[out=0,in=180] (0,0.5) to[out=0,in=180] (B.center);
			\begin{scope}[shift={(-3,0)}]
				\begin{scope}[scale=0.3]
					\foreach \x/\y in {0/0, 0/1, 0/2, 0/3, 0/4, 1/0, 1/1, 1/2, 2/0, 2/1, 2/2, 3/0, 3/1, 3/2, 4/0}
					{
						\draw[fill=white] (0.866025*\x, -0.5*\x + \y) -- (0.866025*\x, -0.5*\x + \y + 1) -- (0.866025*\x + 0.866025, -0.5*\x + \y + 0.5) -- (0.866025*\x + 0.866025, -0.5*\x + \y -0.5) -- cycle;
					}
					\foreach \x/\y in {1/3}
					{
						\draw[thick, dashed, \myblue, fill=white] (0.866025*\x, -0.5*\x + \y) -- (0.866025*\x, -0.5*\x + \y + 1) -- (0.866025*\x + 0.866025, -0.5*\x + \y + 0.5) -- (0.866025*\x + 0.866025, -0.5*\x + \y -0.5) -- cycle;
					}
				\end{scope}
			\end{scope}
			\begin{scope}[shift={(0,0.5)}]
				\begin{scope}[scale=0.3]
					\foreach \x/\y in {0/0}
					{
						\draw[fill=\myblue] (0.866025*\x, -0.5*\x + \y) -- (0.866025*\x, -0.5*\x + \y + 1) -- (0.866025*\x + 0.866025, -0.5*\x + \y + 0.5) -- (0.866025*\x + 0.866025, -0.5*\x + \y -0.5) -- cycle;
					}
				\end{scope}
			\end{scope}
			\node[below left] at (-3,0) {$\scriptstyle m_1$};
			\node[below left] at (3,0) {$\scriptstyle m_2$};
			\node[above left] at (-3,0) {$\scriptstyle \Kappa_1$};
			\node[above right] at (3.7,0) {$\scriptstyle \Kappa_2$};
			\begin{scope}[shift={(0,-2)}]
				\draw[postaction={decorate}, decoration={markings, mark= at position 0.65 with {\arrow{stealth}}}] (0,-1) to[out=120,in=240] (0,0);
				\draw[postaction={decorate}, decoration={markings, mark= at position 0.65 with {\arrow{stealth}}}] (0,-1) to[out=60,in=300] (0,0);
				\begin{scope}[shift = {(0, -1)}]
					\draw[fill=\myblue] (-0.08,-0.08) -- (-0.08,0.08) -- (0.08,0.08) -- (0.08,-0.08) -- cycle;
				\end{scope}
				\foreach \x/\y in {-0.110854/-0.438772, -0.0397871/-0.414202, 0.175573/0.0123692, 0.0907203/0.349026, 0.054705/0.326349, -0.445493/-0.153395, -0.310788/0.474014, 0.333262/-0.081624, -0.140032/-0.411218, 0.296149/0.493719, -0.479132/-0.416032, 0.27214/0.46536, 0.131722/-0.477259, -0.188073/0.379562, 0.456149/0.0103718, 0.221207/-0.469464, -0.0985561/0.184023, 0.1667/0.183931, -0.287768/0.210008, 0.333437/-0.234445}
				{
					\draw[thin, dashed] (0.8*\x-3., 0.8*\y) -- (0,0);
					\begin{scope}[shift = {(0.8*\x-3., 0.8*\y)}]
						\draw[fill=white] (-0.06,-0.06) -- (-0.06,0.06) -- (0.06,0.06) -- (0.06,-0.06) -- cycle;
					\end{scope}
				}
				\foreach \x/\y in {0.246682/0.182191, -0.426704/-0.304155, -0.317677/-0.0175759, 0.267578/-0.208248, 0.483216/0.170115, -0.0385972/-0.140475, -0.0268149/0.0490305, 0.0469072/0.307521, -0.278007/0.0567119, -0.206365/0.0364349, -0.289147/0.41706, 0.430238/0.136712, -0.0358284/-0.265131, 0.356943/-0.0591335, -0.218152/0.252445, -0.410636/-0.350886, -0.201367/-0.41767, 0.127961/0.289589, 0.325448/0.0332991, -0.418946/0.482068}
				{
					\draw[thin, dashed] (0.8*\x+3., 0.8*\y) -- (0,0);
					\begin{scope}[shift = {(0.8*\x+3., 0.8*\y)}]
						\draw[fill=white] (-0.06,-0.06) -- (-0.06,0.06) -- (0.06,0.06) -- (0.06,-0.06) -- cycle;
					\end{scope}
				}
				\draw[fill=\myblue] (0,0) circle (0.1);
				\draw[gray, dashed, thick] (-0.3,0.3) -- (0.3,0.3) -- (0.3,-1.3) -- (-0.3,-1.3) -- cycle;
				\node[gray, right] at (0.3,-0.5) {$\IC\IP^1$};
			\end{scope}
			\draw[->] (-3,-0.3) -- (-3,-1.4);
			\draw[->] (0,-0.3) -- (0,-1.4);
			\draw[->] (3,-0.3) -- (3,-1.4);
		\end{tikzpicture}
	\end{center}
	\caption{IR quiver description of an atom migration.}\label{fig:migration}
\end{figure}

Consider a migration of a single atom $\Box$ from crystal $\Kappa_1$ to crystal $\Kappa_2$.
Effectively we could approximate this situation as steady atoms of $\Kappa_1$ and $\Kappa_2$ are ``frozen'', so that a migration of a single $\sigma$ could be pictured as an effective IR transition in a $U(1)$ theory where frozen fields become flavor symmetries, where $\sigma$ is a complex scalar in the $U(1)$ gauge multiplet (see fig.\ref{fig:migration}).

Having this in mind we consider a simple toy model, where in addition to a $\IC\IP^1$ base there are only pairs of positively and negatively $U(1)$-charged fiber fields mimicking frozen crystal degrees of freedom.
\begin{equation}
	\begin{array}{c}
		\begin{tikzpicture}
			\draw[postaction={decorate},
			decoration={markings, mark= at position 0.65 with {\arrow{stealth}}}] (-2,-0.3) -- (0,0) node[pos=0.3,below] {$\scriptstyle\phi_{2+}$};
			\draw[postaction={decorate},
			decoration={markings, mark= at position 0.65 with {\arrow{stealth}}}] (-2,0.3) -- (0,0) node[pos=0.3,above] {$\scriptstyle\phi_{1+}$};
			\draw[postaction={decorate},
			decoration={markings, mark= at position 0.65 with {\arrow{stealth}}}] (0,0) -- (2,-0.3) node[pos=0.7,below] {$\scriptstyle\phi_{2-}$};
			\draw[postaction={decorate},
			decoration={markings, mark= at position 0.65 with {\arrow{stealth}}}] (0,0) -- (2,0.3) node[pos=0.7,above] {$\scriptstyle\phi_{1-}$};
			\draw[postaction={decorate},
			decoration={markings, mark= at position 0.65 with {\arrow{stealth}}}] (-0.2,-1.5) -- (0,0) node[pos=0.5,left] {$\scriptstyle X_1$};
			\draw[postaction={decorate},
			decoration={markings, mark= at position 0.65 with {\arrow{stealth}}}] (0.2,-1.5) -- (0,0) node[pos=0.5,right] {$\scriptstyle X_2$};
			\draw[fill=white] (0,0) circle (0.1);
			\node[above] at (0,0.1) {$\scriptstyle\sigma$};
			\begin{scope}[shift={(-2,-0.3)}]
				\draw[fill=white] (-0.08,-0.08) -- (-0.08,0.08) -- (0.08,0.08) -- (0.08,-0.08) -- cycle;
				\node[left] at (-0.08,0) {$\scriptstyle (\mu_{2+},m_2)$};
			\end{scope}
			\begin{scope}[shift={(-2,0.3)}]
				\draw[fill=white] (-0.08,-0.08) -- (-0.08,0.08) -- (0.08,0.08) -- (0.08,-0.08) -- cycle;
				\node[left] at (-0.08,0) {$\scriptstyle (\mu_{1+},m_1)$};
			\end{scope}
			\begin{scope}[shift={(2,-0.3)}]
				\draw[fill=white] (-0.08,-0.08) -- (-0.08,0.08) -- (0.08,0.08) -- (0.08,-0.08) -- cycle;
				\node[right] at (0.08,0) {$\scriptstyle (\mu_{2-},m_2)$};
			\end{scope}
			\begin{scope}[shift={(2,0.3)}]
				\draw[fill=white] (-0.08,-0.08) -- (-0.08,0.08) -- (0.08,0.08) -- (0.08,-0.08) -- cycle;
				\node[right] at (0.08,0) {$\scriptstyle (\mu_{1-},m_1)$};
			\end{scope}
			\begin{scope}[shift={(-0.2,-1.5)}]
				\draw[fill=white] (-0.08,-0.08) -- (-0.08,0.08) -- (0.08,0.08) -- (0.08,-0.08) -- cycle;
				\node[left] at (-0.08,0) {$\scriptstyle (a_1,m_1)$};
			\end{scope}
			\begin{scope}[shift={(0.2,-1.5)}]
				\draw[fill=white] (-0.08,-0.08) -- (-0.08,0.08) -- (0.08,0.08) -- (0.08,-0.08) -- cycle;
				\node[right] at (0.08,0) {$\scriptstyle (a_2,m_2)$};
			\end{scope}
		\end{tikzpicture}
	\end{array}\cong
	\left(\begin{array}{c}
		\begin{tikzpicture}
			\draw[postaction={decorate},
			decoration={markings, mark= at position 0.65 with {\arrow{stealth}}}] (-1,0) -- (0,0) node[pos=0.5,above] {$\scriptstyle \phi_{1+}$};
			\draw[postaction={decorate},
			decoration={markings, mark= at position 0.65 with {\arrow{stealth}}}] (0,0) -- (1,0) node[pos=0.5,above] {$\scriptstyle \phi_{1-}$};
			\draw[postaction={decorate},
			decoration={markings, mark= at position 0.65 with {\arrow{stealth}}}] (0,-1) -- (0,0) node[pos=0.3,right] {$\scriptstyle X_1$};
			\draw[fill=white] (0,0) circle (0.1);
			\begin{scope}[shift={(-1,0)}]
				\draw[fill=white] (-0.08,-0.08) -- (-0.08,0.08) -- (0.08,0.08) -- (0.08,-0.08) -- cycle;
				\node[below] at (0,-0.08) {$\scriptstyle \mu_{1+}$};
			\end{scope}
			\begin{scope}[shift={(1,0)}]
				\draw[fill=white] (-0.08,-0.08) -- (-0.08,0.08) -- (0.08,0.08) -- (0.08,-0.08) -- cycle;
				\node[below] at (0,-0.08) {$\scriptstyle \mu_{1-}$};
			\end{scope}
			\begin{scope}[shift={(0,-1)}]
				\draw[fill=white] (-0.08,-0.08) -- (-0.08,0.08) -- (0.08,0.08) -- (0.08,-0.08) -- cycle;
				\node[below] at (0,-0.08) {$\scriptstyle a_1$};
			\end{scope}
		\end{tikzpicture}
	\end{array}\right)_{m_1}
	\otimes
	\left(\begin{array}{c}
		\begin{tikzpicture}
			\draw[postaction={decorate},
			decoration={markings, mark= at position 0.65 with {\arrow{stealth}}}] (-1,0) -- (0,0) node[pos=0.5,above] {$\scriptstyle \phi_{2+}$};
			\draw[postaction={decorate},
			decoration={markings, mark= at position 0.65 with {\arrow{stealth}}}] (0,0) -- (1,0) node[pos=0.5,above] {$\scriptstyle \phi_{2-}$};
			\draw[postaction={decorate},
			decoration={markings, mark= at position 0.65 with {\arrow{stealth}}}] (0,-1) -- (0,0) node[pos=0.3,right] {$\scriptstyle X_2$};
			\draw[fill=white] (0,0) circle (0.1);
			\begin{scope}[shift={(-1,0)}]
				\draw[fill=white] (-0.08,-0.08) -- (-0.08,0.08) -- (0.08,0.08) -- (0.08,-0.08) -- cycle;
				\node[below] at (0,-0.08) {$\scriptstyle \mu_{2+}$};
			\end{scope}
			\begin{scope}[shift={(1,0)}]
				\draw[fill=white] (-0.08,-0.08) -- (-0.08,0.08) -- (0.08,0.08) -- (0.08,-0.08) -- cycle;
				\node[below] at (0,-0.08) {$\scriptstyle \mu_{2-}$};
			\end{scope}
			\begin{scope}[shift={(0,-1)}]
				\draw[fill=white] (-0.08,-0.08) -- (-0.08,0.08) -- (0.08,0.08) -- (0.08,-0.08) -- cycle;
				\node[below] at (0,-0.08) {$\scriptstyle a_2$};
			\end{scope}
		\end{tikzpicture}
	\end{array}\right)_{m_2}\,.
\end{equation}

The height function \eqref{height} reads in this case:
\begin{equation}
	U=hr-(h-m_1)|X_1|^2-(h-m_2)|X_2|^2\,.
\end{equation}

Let us change coordinates ($\varphi,\vartheta\in[0,2\pi)]$, $\chi\in[0,\pi/2]$):
\begin{equation}
	X_1=\rho e^{\I\varphi}\cos\chi,\quad X_2=\rho e^{\I\varphi}\sin\chi e^{\I\vartheta}\,.
\end{equation}

In these terms the height function reads ($m(\chi)=m_1\cos^2\chi+m_2\sin^2\chi$):
\begin{equation}
	U=(h-m(\chi))(r-\rho^2)+\frac{m_{12}}{2}r\cos2\chi\,.
\end{equation}
The first term in the height function forces the effective theory to live on a 3-sphere of radius $\rho=\sqrt{r}$ (a Hopf fibration over the $\IC\IP^1$ base parameterized by $X_1:X_2$) and fixes the expectation value for $h=m(\chi)$.
The second term delivers two cassical vacua -- south and north pole of $\IC\IP^1$:
\begin{equation}
	\begin{array}{lll}
		\chi=0,&\cos2\chi=+1, & (\Box,\varnothing)\,;\\
		\chi=\frac{\pi}{2},&\cos2\chi=-1, & (\varnothing,\Box)\,.\\
	\end{array}
\end{equation}
corresponding to the situations when the atom resides in $\Kappa_1$ or $\Kappa_2$ respectively in the initial model.
This term induces instantonic tunneling between vacua satisfying \eqref{instanton}:
\begin{equation}
	\cos\chi(\tau)=\frac{1}{\sqrt{1+e^{8m_{21}\tau}}},\quad \sin\chi(\tau)=\frac{e^{4m_{21}\tau}}{\sqrt{1+e^{8m_{21}\tau}}}\,,
\end{equation}
where $\tau$ is the Euclidean time along the instanton trajectory.

Thus if $m_2>m_1$ then $Q_1$-instanton flows form vacuum $(\Box,\varnothing)$ to vacuum $(\varnothing,\Box)$ as the Morse height function $U$ increases only in this direction.

The resulting wave function for IR wave function could be written in the following form for two vacua as a factorization of elementary wave functions for free chiral scalars:
\begin{equation}
	\begin{split}
		\Psi_{(\Box,\varnothing)}=&\Phi\left(\phi_{1+}|a_1-\mu_{1+},m_1-m_1\right)\Phi\left(\phi_{1-}|\mu_{1-}-a_1,m_1-m_1\right)\times\\
		\times&\Phi\left(\phi_{2+}|a_1-\mu_{2+},m_1-m_2\right)\Phi\left(\phi_{2-}|\mu_{2-}-a_1,m_2-m_1\right)\Phi(X_1:X_2|a_1-a_2,m_1-m_2),\\
		\Psi_{(\varnothing,\Box)}=&\Phi\left(\phi_{1+}|a_2-\mu_{1+},m_2-m_1\right)\Phi\left(\phi_{1-}|\mu_{1-}-a_2,m_1-m_2\right)\times\\
		\times&\Phi\left(\phi_{2+}|a_2-\mu_{2+},m_2-m_2\right)\Phi\left(\phi_{2-}|\mu_{2-}-a_2,m_2-m_2\right)\Phi(X_1:X_2|a_2-a_1,m_2-m_1),\\
	\end{split}
\end{equation}
where $X_1:X_2$ is a homogeneous coordinate on the tangent fiber to $\IC\IP^1$.
Function $\Phi(\phi|z,x_3)$ represents a wave function of an equivariant $\IC$-plane discussed in Appendix \ref{app:Berry}.

All the weight parameters are effective and vary along the instanton trajectory as $h=m(\chi)$ depends on the $\chi$-value effectively.
Thus we conclude that the instanton induces flows on the moduli space for these fields and respective Berry connection transport.
In our model for the connection not only a mode of weight $\sigma$ contributes, rather all modes $\sigma+p$ with momenta $p$ along $\Sigma$ contribute.
According to regularization scheme \eqref{regularization} this leads to a substitution of each weight space Euler class $\sigma$ by a Conner-Floyd class ${\bf e}_{\xi}(\sigma)$.
In the following table we calculate transports for fields following \eqref{holo_transport}.
\begin{equation}\label{amps}
	\begin{array}{c|c|c}
		\mbox{Field} & \mbox{Flow direction} & \mbox{Amplitude} \\
		\hline
		(1+) & \mbox{equator}\to\mbox{north} & 1\\
		(1-) & \mbox{equator}\to\mbox{south} & {\bf e}_{\xi}\left(T\phi_{(1-)}\right)\\
		(2+) & \mbox{south}\to\mbox{equator} & {\bf e}_{\xi}\left(T\phi_{(2+)}\right)^{-1}\\
		(2-) & \mbox{north}\to\mbox{equator} & 1\\
		\IC\IP^1 & \mbox{south}\to\mbox{north} &{\bf e}_{\xi}\left(T(X_1:X_2)\right)^{-1}\\
	\end{array}
\end{equation}
where $T\phi$ denotes the tangent space to the complex plane spanned by the corresponding chiral field $\phi$.

The total amplitude is a classical R-matrix for the associated quiver BPS algebra in a form of a t-channel amplitude (the instanton Euclidean time $\tau$ flows upwards in the diagram, compare to \cite[sec.~4.4.5]{Galakhov:2022uyu}):
\begin{equation}\label{instanton_amp}
	\begin{array}{c}
		\begin{tikzpicture}
			\node(A) at (-4,0) {${\bf e}_{\xi}\left(T\phi_{(1-)}\right)$};
			\node(B) at (4,0) {${\bf e}_{\xi}\left(T\phi_{(2+)}\right)^{-1}$};
			\node(C) at (0,0) {${\bf e}_{\xi}\left(T(X_1:X_2)\right)^{-1}$};
			\node at (-2.3,0) {$\cdot$};
			\node at (2.3,0) {$\cdot$};
			\node[left] at (A.west) {$\mathscr{A}=$};
			\node[right] at (B.east) {;};
			\begin{scope}[shift={(0,-2)}]
				\draw[thick,\myblue,decorate,decoration={snake,amplitude=3pt, segment length=20pt}] (-2,0) -- (2,0);
				\draw[ultra thick, postaction={decorate},
				decoration={markings, mark= at position 0.65 with {\arrow{stealth}}}] (-2.5,-1) -- (-2,0);
				\draw[ultra thick, postaction={decorate},
				decoration={markings, mark= at position 0.65 with {\arrow{stealth}}}] (-2,0) -- (-2.5,1);
				\draw[ultra thick, postaction={decorate},
				decoration={markings, mark= at position 0.65 with {\arrow{stealth}}}] (2.5,-1) -- (2,0);
				\draw[ultra thick, postaction={decorate},
				decoration={markings, mark= at position 0.65 with {\arrow{stealth}}}](2,0) -- (2.5,1);
				\node[left] at (-2.5,1) {$\Kappa_1-\Box$};
				\node[left] at (-2.5,-1) {$\Kappa_1$};
				\node[right] at (2.5,1) {$\Kappa_2+\Box$};
				\node[right] at (2.5,-1) {$\Kappa_2$};
				\draw[fill=orange] (-2,0) circle (0.08) (2,0) circle (0.08);
			\end{scope}
			\draw[->] (0,-0.4) -- (0,-1.6) node[pos=0.5, right]{\tiny propagator};
			\draw[->] (-4,-0.4) to[out=270,in=90] (-5,-1.4) to[out=270,in=180] (-2.3,-2);
			\node[left] at (-5,-1.4) {$\begin{array}{c}
					\mbox{\tiny lowering vertex}\\
					\scriptstyle \left[\Kappa_1\to\Kappa_1-\Box\right]
				\end{array}$};
			\draw[->] (4,-0.4) to[out=270,in=90] (5,-1.4) to[out=270,in=0] (2.3,-2);
			\node[right] at (5,-1.4) {$\begin{array}{c}
					\mbox{\tiny raising vertex}\\
					\scriptstyle \left[\Kappa_2\to\Kappa_2+\Box\right]
				\end{array}$};
		\end{tikzpicture}
	\end{array}
\end{equation}

In general, we would like to argue that the instanton amplitude could be split in the following way:
\begin{equation}
	\mathscr{A}=\left[\Kappa_1\to\Kappa_1-\Box\right]\cdot{\bf e}_{\xi}\left(\sigma_{\rm fin}-\sigma_{\rm in}\right)^{-1}\cdot \left[\Kappa_2\to\Kappa_2+\Box\right]\,,
\end{equation}
where $\sigma_{\rm in}$ and $\sigma_{\rm fin}$ are initial and final positions of the migrating atom in the $\sigma$-plane in the initial crystal and the final one respectively.
Vertices corresponding to atom ``emission'' and ``capture'' processes turn out to coincide with ones defined in \eqref{matrix}.

Let us check how this prescription works in our effective model.
Incidence loci are calculated from the following commutative diagrams ($\alpha=1,2$):
\begin{equation}
	\begin{array}{c}
		\begin{tikzpicture}
			\node (A) at (0,0) {$\IC_{\mu_{\alpha+}}$};
			\node (B) at (2,0) {$\IC_{a_{\alpha}}$};
			\node (C) at (4,0) {$\IC_{\mu_{\alpha-}}$};
			\node (D) at (0,-1.5) {$\IC_{\mu_{\alpha+}}$};
			\node (E) at (2,-1.5) {$0$};
			\node (F) at (4,-1.5) {$\IC_{\mu_{\alpha-}}$};
			\path (A) edge[->] node[above] {$\scriptstyle\phi_{\alpha+}$} (B) (B) edge[->] node[above] {$\scriptstyle\phi_{\alpha-}$} (C) (D) edge[->] node[above] {$\scriptstyle 0$} (E) (E) edge[->] node[above] {$\scriptstyle 0$} (F) (A) edge[->] node[left] {$\scriptstyle \rm id$} (D) (B) edge[->] node[right] {$\scriptstyle 0$} (E) (C) edge[->] node[right] {$\scriptstyle \rm id$} (F);
		\end{tikzpicture}
	\end{array}\,.
\end{equation}
We have the following tangent spaces:
\begin{equation}
	T\CR_{\Box_{\alpha}}={\rm Span}\left\{\phi_{\alpha+},\phi_{\alpha-}\right\},\quad T\CR_{\varnothing_{\alpha}}=0,\quad T\CI_{\varnothing_{\alpha},\Box_{\alpha}}={\rm Span}\left\{\phi_{\alpha+}\right\}\,.
\end{equation}

Resulting raising/lowering vertex contributions calculated via prescription \eqref{matrix} coincide with ones calculated via the effective Berry connection in \eqref{instanton_amp}:
\begin{equation}
	\begin{split}
		&[\Box_1\to\varnothing_1]=\frac{{\bf e}_{\xi}\left(T\CR_{\Box_1}\right)}{{\bf e}_{\xi}\left(T\CI_{\varnothing_{1},\Box_{1}}\right)}=\frac{{\bf e}_{\xi}\left(T\phi_{(1+)}\right){\bf e}_{\xi}\left(T\phi_{(1-)}\right)}{{\bf e}_{\xi}\left(T\phi_{(1+)}\right)}={\bf e}_{\xi}\left(T\phi_{(1-)}\right)\,,\\
		&[\varnothing_2\to\Box_2]=\frac{{\bf e}_{\xi}\left(T\CR_{\varnothing_2}\right)}{{\bf e}_{\xi}\left(T\CI_{\varnothing_{2},\Box_{2}}\right)}={\bf e}_{\xi}\left(T\phi_{(2+)}\right)^{-1}\,.
	\end{split}
\end{equation}

Our derivation based on the interface Berry connection of the classical R-matrix contribution to the quantum R-matrix is similar to a derivation of \emph{elliptic} stable envelopes \cite{Dedushenko:2021mds,Bullimore:2021rnr} constituting the quantum R-matrix \cite{2012arXiv1211.1287M,Aganagic:2016jmx} (cf. elementary transport amplitudes for chirals \eqref{amps} and \cite[sec.~3.3]{Bullimore:2021rnr}).


\subsection{BPS algebra from a physical viewpoint}

As we have noted the structure of the instanton amplitude suggests a diagrammatic expansion in the t-channel \eqref{instanton_amp}: two vertices and a propagator.
We have called two vertices ``lowering'' and ``raising'' as schematically we could represent these vertices as elementary processes of emanating or capturing of an atom by a crystal.
We treat these vertex elements as matrix elements of an algebra of operators adding/subtracting atoms to/from crystals so that result is a crystal again.
Similarly to \cite[sec.~2.7]{Galakhov:2020vyb} we treat the origin of expressions for these matrix coefficients as transport with respect to the induced Berry connection.
Before calculating relations between these matrix elements let us introduce a basic element of Berry transport we will need in what follows as a permutation of two atoms of colors $i$ and $j$ along the $h$-axis.

The permutation occurs in the ordering of atom positions along the $h$-axis whereas their positions in the $\sigma$-plane is preserved.
The result depends only on a relative position in the $\sigma$-plane we denote simply as $\sigma_{12}=-\sigma_{21}$.
We have already calculated that the Berry connection contributes as a form-factor in the CoHA multiplication formula \eqref{eta} for a specific ordering on the $h$-axis.
The permutation of atoms on the $h$-axis contributes simply as a ratio of two factors \eqref{eta} with permuted parameters:
\begin{equation}\label{Prm}
	{\rm Prm}(\sqbox{$i$},\sqbox{$j$})=\frac{\eta_{i,j}(\sigma_{21})}{\eta_{j,i}(\sigma_{12})}=\left(\frac{{\bf e}_{\xi}(\sigma_{12})}{{\bf e}_{\xi}(-\sigma_{12})}\right)^{\delta_{i,j}}
		\frac{\prod\lm_{(a:i\to j)\in \fQ_1}{\bf e}_{\xi}(-\sigma_{12}-\mu_{a})}{\prod\lm_{(b:j\to i)\in\fQ_1}{\bf e}_{\xi}(\sigma_{12}-\mu_{b})}=:\varphi_{i,j}(\sigma_{12})\,.
\end{equation}
As we will see in what follows this is a quantity defining the BPS algebra in this setting.
Note that it is defined completely by the quiver diagram and flavor parameters.
In \cite{Li:2020rij} it acquired a name \emph{``bond factor''} as a quiver depiction is reminiscent of a molecule depiction in chemistry.

We consider the BPS algebra as acting on the space of crystals $\Kappa$ by adding/subtracting atoms $\Kappa\to \Kappa\pm\Box$.
We identify corresponding matrix elements with the vertices in the diagram technique discussed in the previous subsection:
\begin{equation}
	[\Kappa\to \Kappa\pm\Box]\,.
\end{equation}
In what follows we will consider non-trivial paths in the crystal space and denote corresponding transition matrix elements accordingly:
\begin{equation}
	\left[\Kappa_1\to\Kappa_2\to\Kappa_3\right]:=\left[\Kappa_1\to\Kappa_2\right]\cdot \left[\Kappa_2\to\Kappa_3\right]\,.
\end{equation}

Consider first a relation emerging when we permute the action of two raising generators.
We could arrive from some crystal $\Kappa$ to a crystal $\Kappa+\Box_1+\Box_2$ in two ways: first we add atom $\Box_1$ to its position then atom $\Box_2$ or vise versa.
Hysteresis between these two paths in the crystal space boils down to a permutation of ordering atoms $\Box_1$ and $\Box_2$ in the $h$-plane as the instanton raising vertex adds these atoms in a different order in two processes:
\begin{equation}
	\begin{array}{c}
		\begin{tikzpicture}[rotate=120]
			\tikzset{cb1/.style = {fill=black!40!red}}
			\tikzset{cb2/.style = {fill=black!40!blue}}
			\tikzset{cl1/.style = {black!40!red}}
			\tikzset{cl2/.style = {black!40!blue}}
			\begin{scope}[scale=0.7]
				\draw[thick, -stealth] (0,0) -- (0.866025, -0.5);
				\draw[thick, -stealth] (0,0) -- (-0.866025, -0.5);
			\end{scope}
			\draw[thick,-stealth] (0,0) -- (0,3);
			\begin{scope}[shift={(0,2.5)}]
				\begin{scope}[scale=0.27]
					\foreach \a/\b in {1/3}
					{
						\node (A) at (0.866025*\a -0.866025*\b , -0.5*\a-0.5*\b) {};
					}
					\foreach \a/\b in {3/2}
					{
						\node (B) at (0.866025*\a -0.866025*\b , -0.5*\a-0.5*\b) {};
					}
				\end{scope}
			\end{scope}
			\draw[cl1, thick, postaction={decorate},
			decoration={markings, mark= at position 0.65 with {\arrow{stealth}}}] (0,0.7) to[out=90,in=270] (A.center);
			\draw[cb1] (0,0.7) circle (0.05);
			\node[below, cl1] at (0,0.7) {1};
			\draw[cl2, thick,postaction={decorate},
			decoration={markings, mark= at position 0.65 with {\arrow{stealth}}}] (0,0) to[out=70,in=270] (B.center);
			\draw[cb2] (0,0) circle (0.05);
			\node[above right, cl2] at (0,0) {2};
			\begin{scope}[shift={(0,2.5)}]
				\begin{scope}[scale=0.27]
					\foreach \a/\b in {0/0,0/1,0/2,0/3, 1/0,1/1,1/2, 2/0,2/1,2/2, 3/0,3/1, 4/0,4/1, 5/0}
					{
						\draw[fill=white] (0.866025*\a -0.866025*\b , -0.5*\a-0.5*\b) -- (0.866025*\a -0.866025*\b + 0.866025, -0.5*\a-0.5*\b-0.5) -- (0.866025*\a -0.866025*\b, -0.5*\a-0.5*\b-1) -- (0.866025*\a -0.866025*\b -0.866025, -0.5*\a-0.5*\b -0.5) -- cycle;
					}
					\foreach \a/\b in {1/3}
					{
						\draw[cb1] (0.866025*\a -0.866025*\b , -0.5*\a-0.5*\b) -- (0.866025*\a -0.866025*\b + 0.866025, -0.5*\a-0.5*\b-0.5) -- (0.866025*\a -0.866025*\b, -0.5*\a-0.5*\b-1) -- (0.866025*\a -0.866025*\b -0.866025, -0.5*\a-0.5*\b -0.5) -- cycle;
						\node[white] at (0.866025*\a -0.866025*\b , -0.5*\a-0.5*\b-0.5) {\tiny 1};
					}
					\foreach \a/\b in {3/2}
					{
						\draw[cb2] (0.866025*\a -0.866025*\b , -0.5*\a-0.5*\b) -- (0.866025*\a -0.866025*\b + 0.866025, -0.5*\a-0.5*\b-0.5) -- (0.866025*\a -0.866025*\b, -0.5*\a-0.5*\b-1) -- (0.866025*\a -0.866025*\b -0.866025, -0.5*\a-0.5*\b -0.5) -- cycle;
						\node[white] at (0.866025*\a -0.866025*\b , -0.5*\a-0.5*\b-0.5) {\tiny 2};
					}
				\end{scope}
			\end{scope}
		\end{tikzpicture}
	\end{array}
	\quad\Bigg/\quad
	\begin{array}{c}
		\begin{tikzpicture}[rotate=120]
			\tikzset{cb1/.style = {fill=black!40!red}}
			\tikzset{cb2/.style = {fill=black!40!blue}}
			\tikzset{cl1/.style = {black!40!red}}
			\tikzset{cl2/.style = {black!40!blue}}
			\begin{scope}[scale=0.7]
				\draw[thick, -stealth] (0,0) -- (0.866025, -0.5);
				\draw[thick, -stealth] (0,0) -- (-0.866025, -0.5);
			\end{scope}
			\draw[thick,-stealth] (0,0) -- (0,3);
			\begin{scope}[shift={(0,2.5)}]
				\begin{scope}[scale=0.27]
					\foreach \a/\b in {1/3}
					{
						\node (A) at (0.866025*\a -0.866025*\b , -0.5*\a-0.5*\b) {};
					}
					\foreach \a/\b in {3/2}
					{
						\node (B) at (0.866025*\a -0.866025*\b , -0.5*\a-0.5*\b) {};
					}
				\end{scope}
			\end{scope}
			\draw[cl1, thick, postaction={decorate},
			decoration={markings, mark= at position 0.65 with {\arrow{stealth}}}] (0,0) to[out=90,in=270] (A.center);
			\draw[cb1] (0,0) circle (0.05);
			\node[below,cl1] at (0,0) {1};
			\draw[cl2, thick,postaction={decorate},
			decoration={markings, mark= at position 0.65 with {\arrow{stealth}}}] (0,0.7) to[out=90,in=270] (B.center);
			\draw[cb2] (0,0.7) circle (0.05);
			\node[above,cl2] at (0,0.7) {2};
			\begin{scope}[shift={(0,2.5)}]
				\begin{scope}[scale=0.27]
					\foreach \a/\b in {0/0,0/1,0/2,0/3, 1/0,1/1,1/2, 2/0,2/1,2/2, 3/0,3/1, 4/0,4/1, 5/0}
					{
						\draw[fill=white] (0.866025*\a -0.866025*\b , -0.5*\a-0.5*\b) -- (0.866025*\a -0.866025*\b + 0.866025, -0.5*\a-0.5*\b-0.5) -- (0.866025*\a -0.866025*\b, -0.5*\a-0.5*\b-1) -- (0.866025*\a -0.866025*\b -0.866025, -0.5*\a-0.5*\b -0.5) -- cycle;
					}
					\foreach \a/\b in {1/3}
					{
						\draw[cb1] (0.866025*\a -0.866025*\b , -0.5*\a-0.5*\b) -- (0.866025*\a -0.866025*\b + 0.866025, -0.5*\a-0.5*\b-0.5) -- (0.866025*\a -0.866025*\b, -0.5*\a-0.5*\b-1) -- (0.866025*\a -0.866025*\b -0.866025, -0.5*\a-0.5*\b -0.5) -- cycle;
						\node[white] at (0.866025*\a -0.866025*\b , -0.5*\a-0.5*\b-0.5) {\tiny 1};
					}
					\foreach \a/\b in {3/2}
					{
						\draw[cb2] (0.866025*\a -0.866025*\b , -0.5*\a-0.5*\b) -- (0.866025*\a -0.866025*\b + 0.866025, -0.5*\a-0.5*\b-0.5) -- (0.866025*\a -0.866025*\b, -0.5*\a-0.5*\b-1) -- (0.866025*\a -0.866025*\b -0.866025, -0.5*\a-0.5*\b -0.5) -- cycle;
						\node[white] at (0.866025*\a -0.866025*\b , -0.5*\a-0.5*\b-0.5) {\tiny 2};
					}
				\end{scope}
			\end{scope}
		\end{tikzpicture}
	\end{array}=\;\;{\rm Prm}({\color{black!40!red} \Box_1},{\color{black!40!blue} \Box_2})\,.
\end{equation}

Translating these pictural descriptions into the introduced language we derive the following relation between matrix elements:
\begin{equation}\label{eq1}
	\frac{[\Kappa\to \Kappa+\sqbox{$a$}\to \Kappa+\sqbox{$a$}+\sqbox{$b$}]}{[\Kappa\to \Kappa+\sqbox{$b$}\to \Kappa+\sqbox{$a$}+\sqbox{$b$}]}=\varphi_{a,b}\left(\omega_{\sqbox{$a$}}-\omega_{\sqbox{$b$}}\right)\,,
\end{equation}
where $\omega_{\sqbox{$a$}}$ and $\omega_{\sqbox{$b$}}$ are positions of atoms $\sqbox{$a$}$ and $\sqbox{$b$}$ in the $z$-plane in the crystals.

Here the ratio of type $A/B=C/D$ as in \eqref{eq1} should be understood as $AD=BC$.
Then by analytic construction \eqref{log} both the log- and exp- functions have a zero at $\sigma=0$, so that ${\bf e}_{\xi}(0)=0$.
This reflects a physical property of \eqref{Prm}.
Whenever a zero or a pole in \eqref{Prm} for given weight values appears one of the paths in the crystal space is unavailable.
For example, path $[\varnothing\to \sqbox{$1$}\to \sqbox{$1$}\!\sqbox{$2$}]$ is available in one variant due to the crystal melting rule \cite[sec.~6.4]{Li:2020rij} meaning that  $[\varnothing\to \sqbox{$2$}\to \sqbox{$1$}\!\sqbox{$2$}]=0$.
On the other hand in this case weights of atoms in the $\sigma$-plane are related as $\omega_{\sqbox{$2$}}=\omega_{\sqbox{$1$}}+\mu_{1\to 2}$ and it contributes by 0 to either numerator or denominator of \eqref{Prm}.

A relation between lowering operators is calculated from a similar hysteresis picture:
\begin{equation}
	\begin{array}{c}
		\begin{tikzpicture}[rotate=120]
			\tikzset{cb1/.style = {fill=black!40!red}}
			\tikzset{cb2/.style = {fill=black!40!blue}}
			\tikzset{cl1/.style = {black!40!red}}
			\tikzset{cl2/.style = {black!40!blue}}
			\draw[thick,-stealth] (0,0) -- (0,2.2);
			\begin{scope}[shift={(0,0)}]
				\begin{scope}[scale=0.27]
					\foreach \a/\b in {0/0, 1/0, 2/0, 3/0, 0/1, 1/1, 2/1,  0/2, 1/2, 2/2, 0/3, 1/3, 0/4, 1/4, 0/5}
					{
						\draw[fill=white] (0.866025*\a -0.866025*\b , -0.5*\a-0.5*\b) -- (0.866025*\a -0.866025*\b + 0.866025, -0.5*\a-0.5*\b-0.5) -- (0.866025*\a -0.866025*\b, -0.5*\a-0.5*\b-1) -- (0.866025*\a -0.866025*\b -0.866025, -0.5*\a-0.5*\b -0.5) -- cycle;
					}
				\end{scope}
				\begin{scope}[scale=0.27]
					\foreach \a/\b in {3/1}
					{
						\draw[cb1] (0.866025*\a -0.866025*\b , -0.5*\a-0.5*\b) -- (0.866025*\a -0.866025*\b + 0.866025, -0.5*\a-0.5*\b-0.5) -- (0.866025*\a -0.866025*\b, -0.5*\a-0.5*\b-1) -- (0.866025*\a -0.866025*\b -0.866025, -0.5*\a-0.5*\b -0.5) -- cycle;
						\node[white] at (0.866025*\a -0.866025*\b , -0.5*\a-0.5*\b-0.5) {\tiny 1};
					}
				\end{scope}
				\begin{scope}[scale=0.27]
					\foreach \a/\b in {2/3}
					{
						\draw[cb2] (0.866025*\a -0.866025*\b , -0.5*\a-0.5*\b) -- (0.866025*\a -0.866025*\b + 0.866025, -0.5*\a-0.5*\b-0.5) -- (0.866025*\a -0.866025*\b, -0.5*\a-0.5*\b-1) -- (0.866025*\a -0.866025*\b -0.866025, -0.5*\a-0.5*\b -0.5) -- cycle;
						\node[white] at (0.866025*\a -0.866025*\b , -0.5*\a-0.5*\b-0.5) {\tiny 2};
					}
				\end{scope}
			\end{scope}
			\begin{scope}[shift={(0,0)}]
				\begin{scope}[scale=0.27]
					\foreach \a/\b in {3/1}
					{
						\node(A) at (0.866025*\a -0.866025*\b , -0.5*\a-0.5*\b) {};
					}
				\end{scope}
				\begin{scope}[scale=0.27]
					\foreach \a/\b in {2/3}
					{
						\node(B) at (0.866025*\a -0.866025*\b , -0.5*\a-0.5*\b) {};
					}
				\end{scope}
			\end{scope}
			\draw[cl1, thick, postaction={decorate},
			decoration={markings, mark= at position 0.65 with {\arrow{stealth}}}] (A.center) to[out=90,in=270] (0,2);
			\draw[cb1] (0,2) circle (0.05);
			\node[above, cl1] at (0,2) {1};
			\draw[cl2, thick, postaction={decorate},
			decoration={markings, mark= at position 0.65 with {\arrow{stealth}}}] (B.center) to[out=90,in=270] (0,1);
			\draw[cb2] (0,1) circle (0.05);
			\node[below, cl2] at (0,1) {2};
		\end{tikzpicture}
	\end{array} \Bigg/ 
	\begin{array}{c}
		\begin{tikzpicture}[rotate=120]
			\tikzset{cb1/.style = {fill=black!40!red}}
			\tikzset{cb2/.style = {fill=black!40!blue}}
			\tikzset{cl1/.style = {black!40!red}}
			\tikzset{cl2/.style = {black!40!blue}}
			\draw[thick,-stealth] (0,0) -- (0,2.2);
			\begin{scope}[shift={(0,0)}]
				\begin{scope}[scale=0.27]
					\foreach \a/\b in {0/0, 1/0, 2/0, 3/0, 0/1, 1/1, 2/1,  0/2, 1/2, 2/2, 0/3, 1/3, 0/4, 1/4, 0/5}
					{
						\draw[fill=white] (0.866025*\a -0.866025*\b , -0.5*\a-0.5*\b) -- (0.866025*\a -0.866025*\b + 0.866025, -0.5*\a-0.5*\b-0.5) -- (0.866025*\a -0.866025*\b, -0.5*\a-0.5*\b-1) -- (0.866025*\a -0.866025*\b -0.866025, -0.5*\a-0.5*\b -0.5) -- cycle;
					}
				\end{scope}
				\begin{scope}[scale=0.27]
					\foreach \a/\b in {3/1}
					{
						\draw[cb1] (0.866025*\a -0.866025*\b , -0.5*\a-0.5*\b) -- (0.866025*\a -0.866025*\b + 0.866025, -0.5*\a-0.5*\b-0.5) -- (0.866025*\a -0.866025*\b, -0.5*\a-0.5*\b-1) -- (0.866025*\a -0.866025*\b -0.866025, -0.5*\a-0.5*\b -0.5) -- cycle;
						\node[white] at (0.866025*\a -0.866025*\b , -0.5*\a-0.5*\b-0.5) {\tiny 1};
					}
				\end{scope}
				\begin{scope}[scale=0.27]
					\foreach \a/\b in {2/3}
					{
						\draw[cb2] (0.866025*\a -0.866025*\b , -0.5*\a-0.5*\b) -- (0.866025*\a -0.866025*\b + 0.866025, -0.5*\a-0.5*\b-0.5) -- (0.866025*\a -0.866025*\b, -0.5*\a-0.5*\b-1) -- (0.866025*\a -0.866025*\b -0.866025, -0.5*\a-0.5*\b -0.5) -- cycle;
						\node[white] at (0.866025*\a -0.866025*\b , -0.5*\a-0.5*\b-0.5) {\tiny 2};
					}
				\end{scope}
			\end{scope}
			\begin{scope}[shift={(0,0)}]
				\begin{scope}[scale=0.27]
					\foreach \a/\b in {3/1}
					{
						\node(A) at (0.866025*\a -0.866025*\b , -0.5*\a-0.5*\b) {};
					}
				\end{scope}
				\begin{scope}[scale=0.27]
					\foreach \a/\b in {2/3}
					{
						\node(B) at (0.866025*\a -0.866025*\b , -0.5*\a-0.5*\b) {};
					}
				\end{scope}
			\end{scope}
			\draw[cl1, thick, postaction={decorate},
			decoration={markings, mark= at position 0.65 with {\arrow{stealth}}}] (A.center) to[out=90,in=270] (0,1);
			\draw[cb1] (0,1) circle (0.05);
			\node[above, cl1] at (0,1) {1};
			\draw[cl2, thick, postaction={decorate},
			decoration={markings, mark= at position 0.65 with {\arrow{stealth}}}] (B.center) to[out=100,in=270] (0,2);
			\draw[cb2] (0,2) circle (0.05);
			\node[below, cl2] at (0,2) {2};
		\end{tikzpicture}
	\end{array} =\;\; {\rm Prm}({\color{black!40!red} \Box_1},{\color{black!40!blue} \Box_2})\,.
\end{equation}

The corresponding relation for matrix coefficients reads:
\begin{equation}\label{eq2}
	\frac{[\Kappa+\sqbox{$a$}+\sqbox{$b$}\to \Kappa+\sqbox{$a$}\to \Kappa]}{[\Kappa+\sqbox{$a$}+\sqbox{$b$}\to \Kappa+\sqbox{$b$}\to \Kappa]}=\varphi_{a,b}\left(\omega_{\sqbox{$a$}}-\omega_{\sqbox{$b$}}\right)\,.
\end{equation}

Relations mixing simultaneously raising and lowering generators may be divided in two types depending if one adds/subtracts two different atoms or the same single atom to/from the crystal.

One-way oriented flow of instantons form lower $U$ to higher $U$ indicates that both trajectories for lowering and raising operators are both directed in the same way.
Therefore there is no difference in the order of adding atom $\Box_1$ and subtracting atom $\Box_2$ if $\Box_1\neq\Box_2$:
\begin{equation}
	\begin{array}{c}
		\begin{tikzpicture}[rotate=120]
			\tikzset{cb1/.style = {fill=black!40!red}}
			\tikzset{cb2/.style = {fill=black!40!blue}}
			\tikzset{cl1/.style = {black!40!red}}
			\tikzset{cl2/.style = {black!40!blue}}
			\begin{scope}[scale=0.7]
				\draw[thick, -stealth] (0,0) -- (0.866025, -0.5);
				\draw[thick, -stealth] (0,0) -- (-0.866025, -0.5);
			\end{scope}
			\draw[thick,-stealth] (0,0) -- (0,3);
			\begin{scope}[shift={(0,1.5)}]
				\begin{scope}[scale=0.27]
					\foreach \a/\b in {2/3}
					{
						\node(B) at (0.866025*\a -0.866025*\b , -0.5*\a-0.5*\b) {};
					}
				\end{scope}
			\end{scope}
			\draw[cl2, thick, postaction={decorate},
			decoration={markings, mark= at position 0.65 with {\arrow{stealth}}}] (0,0) to[out=90,in=270] (B.center);
			\draw[cb2] (0,0) circle (0.05);
			\node[below, cl2] at (0,0) {2};
			\begin{scope}[shift={(0,1.5)}]
				\begin{scope}[scale=0.27]
					\foreach \a/\b in {0/0, 1/0, 2/0, 3/0, 0/1, 1/1, 2/1,  0/2, 1/2, 2/2, 0/3, 1/3, 0/4, 1/4, 0/5}
					{
						\draw[fill=white] (0.866025*\a -0.866025*\b , -0.5*\a-0.5*\b) -- (0.866025*\a -0.866025*\b + 0.866025, -0.5*\a-0.5*\b-0.5) -- (0.866025*\a -0.866025*\b, -0.5*\a-0.5*\b-1) -- (0.866025*\a -0.866025*\b -0.866025, -0.5*\a-0.5*\b -0.5) -- cycle;
					}
				\end{scope}
				\begin{scope}[scale=0.27]
					\foreach \a/\b in {3/1}
					{
						\draw[cb1] (0.866025*\a -0.866025*\b , -0.5*\a-0.5*\b) -- (0.866025*\a -0.866025*\b + 0.866025, -0.5*\a-0.5*\b-0.5) -- (0.866025*\a -0.866025*\b, -0.5*\a-0.5*\b-1) -- (0.866025*\a -0.866025*\b -0.866025, -0.5*\a-0.5*\b -0.5) -- cycle;
						\node[white] at (0.866025*\a -0.866025*\b , -0.5*\a-0.5*\b-0.5) {\tiny 1};
					}
				\end{scope}
				\begin{scope}[scale=0.27]
					\foreach \a/\b in {2/3}
					{
						\draw[cb2] (0.866025*\a -0.866025*\b , -0.5*\a-0.5*\b) -- (0.866025*\a -0.866025*\b + 0.866025, -0.5*\a-0.5*\b-0.5) -- (0.866025*\a -0.866025*\b, -0.5*\a-0.5*\b-1) -- (0.866025*\a -0.866025*\b -0.866025, -0.5*\a-0.5*\b -0.5) -- cycle;
						\node[white] at (0.866025*\a -0.866025*\b , -0.5*\a-0.5*\b-0.5) {\tiny 2};
					}
				\end{scope}
			\end{scope}
			\begin{scope}[shift={(0,1.5)}]
				\begin{scope}[scale=0.27]
					\foreach \a/\b in {3/1}
					{
						\node(A) at (0.866025*\a -0.866025*\b , -0.5*\a-0.5*\b) {};
					}
				\end{scope}
			\end{scope}
			\draw[cl1, thick, postaction={decorate},
			decoration={markings, mark= at position 0.65 with {\arrow{stealth}}}] (A.center) to[out=90,in=270] (0,3);
			\draw[cb1] (0,3) circle (0.05);
			\node[above, cl1] at (0,3) {1};
		\end{tikzpicture}
	\end{array} \;\;=\;\;
	\begin{array}{c}
		\begin{tikzpicture}[rotate=120]
			\tikzset{cb1/.style = {fill=black!40!red}}
			\tikzset{cb2/.style = {fill=black!40!blue}}
			\tikzset{cl1/.style = {black!40!red}}
			\tikzset{cl2/.style = {black!40!blue}}
			\begin{scope}[scale=0.7]
				\draw[thick, -stealth] (0,0) -- (0.866025, -0.5);
				\draw[thick, -stealth] (0,0) -- (-0.866025, -0.5);
			\end{scope}
			\draw[thick,-stealth] (0,0) -- (0,3);
			\begin{scope}[shift={(0,1.5)}]
				\begin{scope}[scale=0.27]
					\foreach \a/\b in {2/3}
					{
						\node(B) at (0.866025*\a -0.866025*\b , -0.5*\a-0.5*\b) {};
					}
				\end{scope}
			\end{scope}
			\draw[cl2, thick, postaction={decorate},
			decoration={markings, mark= at position 0.65 with {\arrow{stealth}}}] (0,0) to[out=90,in=270] (B.center);
			\draw[cb2] (0,0) circle (0.05);
			\node[below, cl2] at (0,0) {2};
			\begin{scope}[shift={(0,1.5)}]
				\begin{scope}[scale=0.27]
					\foreach \a/\b in {0/0, 1/0, 2/0, 3/0, 0/1, 1/1, 2/1,  0/2, 1/2, 2/2, 0/3, 1/3, 0/4, 1/4, 0/5}
					{
						\draw[fill=white] (0.866025*\a -0.866025*\b , -0.5*\a-0.5*\b) -- (0.866025*\a -0.866025*\b + 0.866025, -0.5*\a-0.5*\b-0.5) -- (0.866025*\a -0.866025*\b, -0.5*\a-0.5*\b-1) -- (0.866025*\a -0.866025*\b -0.866025, -0.5*\a-0.5*\b -0.5) -- cycle;
					}
				\end{scope}
				\begin{scope}[scale=0.27]
					\foreach \a/\b in {3/1}
					{
						\draw[cb1] (0.866025*\a -0.866025*\b , -0.5*\a-0.5*\b) -- (0.866025*\a -0.866025*\b + 0.866025, -0.5*\a-0.5*\b-0.5) -- (0.866025*\a -0.866025*\b, -0.5*\a-0.5*\b-1) -- (0.866025*\a -0.866025*\b -0.866025, -0.5*\a-0.5*\b -0.5) -- cycle;
						\node[white] at (0.866025*\a -0.866025*\b , -0.5*\a-0.5*\b-0.5) {\tiny 1};
					}
				\end{scope}
				\begin{scope}[scale=0.27]
					\foreach \a/\b in {2/3}
					{
						\draw[cb2] (0.866025*\a -0.866025*\b , -0.5*\a-0.5*\b) -- (0.866025*\a -0.866025*\b + 0.866025, -0.5*\a-0.5*\b-0.5) -- (0.866025*\a -0.866025*\b, -0.5*\a-0.5*\b-1) -- (0.866025*\a -0.866025*\b -0.866025, -0.5*\a-0.5*\b -0.5) -- cycle;
						\node[white] at (0.866025*\a -0.866025*\b , -0.5*\a-0.5*\b-0.5) {\tiny 2};
					}
				\end{scope}
			\end{scope}
			\begin{scope}[shift={(0,1.5)}]
				\begin{scope}[scale=0.27]
					\foreach \a/\b in {3/1}
					{
						\node(A) at (0.866025*\a -0.866025*\b , -0.5*\a-0.5*\b) {};
					}
				\end{scope}
			\end{scope}
			\draw[cl1, thick, postaction={decorate},
			decoration={markings, mark= at position 0.65 with {\arrow{stealth}}}] (A.center) to[out=90,in=270] (0,3);
			\draw[cb1] (0,3) circle (0.05);
			\node[above, cl1] at (0,3) {1};
		\end{tikzpicture}
	\end{array} \,.
\end{equation}
The relation follows from this pictorial description:
\begin{equation}\label{eq3}
	[\Kappa+\sqbox{$a$}\to \Kappa+\sqbox{$a$}+\sqbox{$b$}\to \Kappa+\sqbox{$b$}]=[\Kappa+\sqbox{$a$}\to \Kappa\to \Kappa+\sqbox{$b$}]
\end{equation}

Finally, adding and subtracting subsequently some atom $\Box$ leads to a permutation of $\Box$ with all the atoms in crystal $\Kappa$:
\begin{equation}\label{sing_atom_braid}
	\begin{array}{c}
		\begin{tikzpicture}[rotate=120]
			\tikzset{cb1/.style = {fill=black!40!red}}
			\tikzset{cb2/.style = {fill=black!40!blue}}
			\tikzset{cl1/.style = {black!40!red}}
			\tikzset{cl2/.style = {black!40!blue}}
			\begin{scope}[scale=0.7]
				\draw[thick, -stealth] (0,0) -- (0.866025, -0.5);
				\draw[thick, -stealth] (0,0) -- (-0.866025, -0.5);
			\end{scope}
			\draw[thick,-stealth] (0,0) -- (0,3);
			\begin{scope}[shift={(0,1.5)}]
				\begin{scope}[scale=0.27]
					\foreach \a/\b in {3/1}
					{
						\node(A) at (0.866025*\a -0.866025*\b , -0.5*\a-0.5*\b) {};
					}
				\end{scope}
			\end{scope}
			\draw[cl1, thick, postaction={decorate},
			decoration={markings, mark= at position 0.65 with {\arrow{stealth}}}] (0,0) to[out=90,in=270] (A.center);
			\draw[cb1] (0,0) circle (0.05);
			\node[below, cl1] at (0,0) {2};
			\begin{scope}[shift={(0,1.5)}]
				\begin{scope}[scale=0.27]
					\foreach \a/\b in {0/0, 1/0, 2/0, 3/0, 0/1, 1/1, 2/1,  0/2, 1/2, 2/2, 0/3, 1/3, 0/4, 1/4, 0/5}
					{
						\draw[fill=white] (0.866025*\a -0.866025*\b , -0.5*\a-0.5*\b) -- (0.866025*\a -0.866025*\b + 0.866025, -0.5*\a-0.5*\b-0.5) -- (0.866025*\a -0.866025*\b, -0.5*\a-0.5*\b-1) -- (0.866025*\a -0.866025*\b -0.866025, -0.5*\a-0.5*\b -0.5) -- cycle;
					}
				\end{scope}
				\begin{scope}[scale=0.27]
					\foreach \a/\b in {3/1}
					{
						\draw[cb1] (0.866025*\a -0.866025*\b , -0.5*\a-0.5*\b) -- (0.866025*\a -0.866025*\b + 0.866025, -0.5*\a-0.5*\b-0.5) -- (0.866025*\a -0.866025*\b, -0.5*\a-0.5*\b-1) -- (0.866025*\a -0.866025*\b -0.866025, -0.5*\a-0.5*\b -0.5) -- cycle;
						\node[white] at (0.866025*\a -0.866025*\b , -0.5*\a-0.5*\b-0.5) {\tiny 1};
					}
				\end{scope}
			\end{scope}
			\draw[cl1, thick, postaction={decorate},
			decoration={markings, mark= at position 0.65 with {\arrow{stealth}}}] (A.center) to[out=90,in=270] (0,3);
			\draw[cb1] (0,3) circle (0.05);
			\node[above, cl1] at (0,3) {1};
		\end{tikzpicture}
	\end{array} \;\;=\;\;{\rm Prm}({\color{black!40!red} \Box_1},\Kappa)\,.
\end{equation}

Let us introduce a phase of braiding a single atom $\sqbox{$i$}$ at position $z$ in the $\sigma$-plane with crystal $\Kappa$:
\begin{equation}
	\Psi_{\Kappa}^{(i)}(z):=\prod\lm_{a:i\to i}{\bf e}_{\xi}(-\mu_a)^{-1}\prod\lm_{\ff\in {\rm flavor}}\varphi_{i,\ff}(z-u_{\ff})\prod\lm_{\sqbox{$j$}\in \fQ_0}\varphi_{i,j}(z-\omega_{\sqbox{$j$}})\,,
\end{equation}
where $\ff$ runs over flavor nodes and $u_{\ff}$ are respective flavor charges.
Naively we would say that \eqref{sing_atom_braid} leads to a relation like:
\begin{equation}\label{fake_eq4}
	[\Kappa\to \Kappa+\sqbox{$i$}\to \Kappa]\sim \Psi^{(i)}_{\Kappa}(\omega_{\sqbox{$i$}})\,,
\end{equation}
however at $z=\omega_{\sqbox{$i$}}$ there is a pole of the charge function $\Psi$ \cite[sec.~6.3]{Li:2020rij}.
Also we note that hysteresis holonomy reasoning does not allow one to restore overall normalization of generators, and therefore the proportionality coefficient in \eqref{fake_eq4}.
The change of normalization factors for $[\Kappa\to\Kappa\pm \Box]$ does not affect relations \eqref{eq1}, \eqref{eq2}, \eqref{eq3} since all of them are homogeneous in the matrix elements.
Using alternative definition \eqref{matrix} for matrix elements of some simple processes one could show that the exact relation reads:
\begin{equation}\label{eq4}
	[\Kappa\to \Kappa+\sqbox{$i$}\to \Kappa]=\lim\lm_{t\to 0}{\bf e}_{\xi}(t)\,\Psi_{\Kappa}^{(i)}(t+ \omega_{\sqbox{$i$}})\,.
\end{equation}

As it was shown in \cite{Galakhov:2021vbo} matrix coefficients defined as \eqref{matrix} satisfy relations \eqref{eq1}, \eqref{eq2}, \eqref{eq3}, \eqref{eq4} we derived in this section.

{\bf Nota bene:} an initial constraint for ${\bf e}_{\xi}(z)$ to be an odd function of $z$ from \cite{Galakhov:2021vbo} may be lifted if a more sophisticated ansatz for the generalized bond factor as in \eqref{Prm} is used (cf.\cite[eq.~(3.23)]{Galakhov:2021vbo}).

After deriving matrix representation for this BPS algebra and the relations between matrix elements one might want to reorganize those elements to derive a canonically looking basis of Chevalley generators.
There is a set of additional BPS operators commuting with the supercharge represented by complex Wilson lines  \cite[sec.~4.2]{Galakhov:2021vbo}.
Combining those generators with the derived BPS algebra we could introduce a representation for Chevalley raising and lowering generators as functions of a spectral parameter $z$:
\begin{equation}\label{gen}
	e^{(i)}(z)|\Kappa\rangle=\sum\lm_{\sqbox{$i$}\in\Kappa^{+}}\frac{\left[\Kappa\to\Kappa+\sqbox{$i$}\right]}{g(z-\omega_{\sqbox{$i$}})}|\Kappa+\sqbox{$i$}\rangle,\quad f^{(i)}(z)|\Kappa\rangle=\sum\lm_{\sqbox{$i$}\in\Kappa^{-}}\frac{\left[\Kappa\to\Kappa-\sqbox{$i$}\right]}{g(z-\omega_{\sqbox{$i$}})}|\Kappa-\sqbox{$i$}\rangle\,,
\end{equation}
where $\Kappa^{\pm}$ are sets of positions in the $\sigma$-plane where an atom could be added/subtracted and the result $\Kappa\pm\Box$ is a crystal again, and $g$ is an arbitrary function having a meaning of a resolvent.
In the cases when we consider ordinary cohomology, K-theory or elliptic cohomology such a function $g$ could be chosen in a way that generators \eqref{gen} together with Cartan generators satisfy a compact set of relations reproducing a rational (default), trigonometric or elliptic generalization of the quiver Yangian \cite{Galakhov:2021vbo}.
However if a similar choice could be made and if the resulting algebra could be refurbished as a set of PBW relations in the Chevalley basis for a generic GCT is an open question.

We comment on the relation between the algebra constructed in this section and the CoHA discussed in sec.~\ref{CoHA} in sec.~\ref{sec:alg_math}.
 

\section*{Acknowledgments}
The author would like to thank Wei Li and Masahito Yamazaki for collaborations on works \cite{Galakhov:2020vyb,Galakhov:2021xum,Galakhov:2021vbo,Galakhov:2022uyu} preceding this paper and fruitful comments on the draft.
Also the author would like to thank Mikhail Kapranov, Nikita Kolganov, Andrei Mironov, Viktor Mishnyakov, Gregory W. Moore and Alexei Morozov for illuminating comments at various stages of this project.
This work is supported by the Russian Science Foundation (Grant No.20-12-00195).


\appendix

\section{Dirac monopole vector-potential as a Berry connection}\label{app:Berry}

Consider a parameter space $\IR^3$ spanned by coordinates $(x_1,x_2,x_3)$.
We introduce a complex structure by choosing a holomorphic coordinate $z=x_1+\I x_2$.

Consider a $\IC$-plane spanned by chiral field $\phi$.
Using parameters we introduce a height function and a complex vector field induced by reparameterizations $\phi\to \bar z\,\phi$:
\begin{equation}
	u=-x_3|\phi|^2,\quad v=\bar z\,\phi\,\p_{\phi}\,.
\end{equation}

Equivariant Dolbeault differential in the Cartan model twisted by the height function reads:
\begin{equation}
	D=e^{-u}\left(d+\iota_{\bar v}\right)e^{u}\,.
\end{equation}

We could promote this expression to a supercharge by associating differential forms to fermions by the following dictionary:
\begin{equation}
	d\phi\rightsquigarrow \chi^{\dagger},\quad d\bar\phi\rightsquigarrow \bar\chi^{\dagger},\quad \iota_{\p/\p\phi}\rightsquigarrow \chi,\quad \iota_{\p/\p\bar\phi}\rightsquigarrow \bar\chi,\quad 1\rightsquigarrow|0\rangle\,.
\end{equation}

This operation gives rise to a nilpotent supercharge and its conjugate:
\begin{equation}
	Q=\chi^{\dagger}\left(\p_{\phi}-x_3\bar\phi\right)+\bar\chi\,z\,\bar\phi,\quad Q^{\dagger}=\chi\left(-\p_{\bar\phi}-x_3\phi\right)+\bar\chi^{\dagger}\,\bar z\,\phi\,.
\end{equation}
The Hamiltonian is given by the Laplacian $H=\left\{Q,Q^{\dagger}\right\}$.

Ground states in such a system are identified with harmonic forms.
Let us distinguish two ground state wave functions we call northern ($\mathscr{N}$) and southern ($\mathscr{S}$):
\begin{equation}\label{states}
	|\mathscr{N}\rangle=\frac{z-(|\vec x|+x_3)\chi^{\dagger}\bar\chi^{\dagger}}{\sqrt{\pi\left(|\vec x|+x_3\right)}}e^{-|\vec x|\cdot|\phi|^2}|0\rangle,\quad |\mathscr{S}\rangle=\frac{(|\vec x|-x_3)-\bar z\,\chi^{\dagger}\bar\chi^{\dagger}}{\sqrt{\pi\left(|\vec x|-x_3\right)}}e^{-|\vec x|\cdot|\phi|^2}|0\rangle\,.
\end{equation}

These wave functions represent the same state -- they belong to the same complex line in the Hilbert space:
\begin{equation}
	\bar z^{\frac{1}{2}}|\mathscr{N}\rangle=z^{\frac{1}{2}}|\mathscr{S}\rangle\,.
\end{equation}

However the Berry connection corresponds to the canonical Dirac monopole vector-potential on the parameter space, and it is smooth either in the northern hemisphere $x_3>0$ or in the southern hemisphere $x_3<0$.
Therefore the names for the wave function expressions:
\begin{equation}
	\begin{split}
		\langle\mathscr{N}|d|\mathscr{N}\rangle&=\frac{1}{4}\left(1-\frac{x_3}{|\vec x|}\right)\left(\frac{dz}{z}-\frac{d\bar z}{\bar z}\right)\,,\\
		\langle\mathscr{S}|d|\mathscr{S}\rangle&=\frac{1}{4}\left(-1-\frac{x_3}{|\vec x|}\right)\left(\frac{dz}{z}-\frac{d\bar z}{\bar z}\right)\,.
	\end{split}
\end{equation}

Consider isospin generator $I_3$:
\begin{equation}
	I_3:=z\p_z-\bar z\p_{\bar z}-\bar\chi^{\dagger}\bar\chi,\quad \left[I_3,Q\right]=\left[I_3,Q^{\dagger}\right]=0,\quad \left[I_3,z\right]=z,\quad\left[I_3,\bar z\right]=-\bar z\,.
\end{equation}

So we could decompose the ground state wave function as a holomorphic in $z$ function and a $I_3$-invariant part:
\begin{equation}\label{I_3}
	|\Psi(\vec x)\rangle=z^{\Theta(x_3)}|I_3\mbox{-inv}\rangle\,,
\end{equation}
where $\Theta$ is the Heaviside step function.

Normalization of states \eqref{states} is natural with respect to the Hermitian norm on the Hilbert space, however if we would like to discuss properties of these states as generators of some cohomology theory a different normalization is in order.
On the equator $x_3=0$ the northern wave function becomes a Thom representative of the Euler class for the equivariant cohomology differential $Q$ in the Cartan model:
\begin{equation}\label{app_euler}
	|\mathscr{N}(x_3=0)\rangle\sim\frac{1}{\sqrt{|\vec x|}}{\bf e}(z)\,.
\end{equation}
so that $\int {\bf e}(z)=1$.
Thus we propose to choose cohomological normalization for the north wave function in the following way:
\begin{equation}
	|\tilde{\mathscr{N}}\rangle:=\sqrt{|\vec x|+x_3}\,|\mathscr{N}\rangle\,.
\end{equation}
Alternatively the south wave function in the regime $|x_3|\gg|z|$ concentrates in the 0-form component and eventually the states shrinks to:
\begin{equation}\label{south_reg}
	|\mathscr{S}(|x_3|\gg|z|)\rangle\sim \sqrt{|\vec x|-x_3}\cdot e^{-|x_3||\phi|^2}\cdot 1\,.
\end{equation}
Apparently this state is annihilated by the supercharge as a Dolbeault differential twisted by the height function $Q\sim e^{-|x_3||\phi|^2}\,\p_{\phi}\, e^{|x_3||\phi|^2}$ in this regime.
The second multiplier in \eqref{south_reg} is exactly due to the height function twist.
So it seems natural to normalize the cohomological state as a simple zero-form 1 before the twist:
\begin{equation}
	|\tilde{\mathscr{S}}\rangle:=\frac{1}{\sqrt{|\vec x|-x_3}}\,|\mathscr{S}\rangle\,.
\end{equation}

In these terms transport between the northern and southern hemispheres becomes holomorphic.
Indeed consider the action of transport $\CT$ from the northern hemisphere to the southern one:
\begin{equation}
	\begin{split}
	&\CT|\tilde{\mathscr{N}}\rangle=\sqrt{|\vec x|+x_3}\,\CT|\mathscr{N}\rangle=\sqrt{|\vec x|+x_3}\,\left(\frac{z}{\bar z}\right)^{\frac{1}{2}}|\mathscr{S}\rangle=\\
	&=\sqrt{|\vec x|+x_3}\cdot\frac{z}{\bar z}\cdot\sqrt{|\vec x|-x_3}\,|\tilde{\mathscr{S}}\rangle=z\cdot|\tilde{\mathscr{S}}\rangle\,.
	\end{split}
\end{equation}
Correspondingly for transport in the opposite direction we have:
\begin{equation}
	\CT^{-1}\cdot|\tilde{\mathscr{S}}\rangle=z^{-1}\cdot |\tilde{\mathscr{N}}\rangle\,.
\end{equation}
If one switches the roles of $Q$ and $Q^{\dagger}$ as differentials by swapping the roles of forms $\chi^{\dagger}$ and vector field $\chi$ the transport would flow in the opposite direction and would be \emph{anti-holomorphic}.

Furthermore from \eqref{app_euler} we have noted that it is natural to extend the north wave function to the equator $x_3=0$, thus we imply that the transport amplitude is accumulated in the segment lying between the equator and the south pole.
We could summarize this information about holomorphic transport in the following table:
\begin{equation}\label{holo_transport}
	\begin{split}
		\mbox{north}\xrightarrow{\hspace*{0.8cm} 1\cdot \hspace*{0.8cm}}\mbox{equator}\xrightarrow{\hspace*{0.8cm} z\cdot \hspace*{0.8cm}}\mbox{south}\,,\\
		\mbox{south}\xrightarrow{\hspace*{0.63cm} z^{-1}\cdot \hspace*{0.63cm}}\mbox{equator}\xrightarrow{\hspace*{0.8cm} 1\cdot \hspace*{0.8cm}}\mbox{north}\,.
	\end{split}
\end{equation}
Note that this transport is equivalent to transport of \eqref{I_3} under an assumption that the $I_3$-invariant part does not vary during this process.



\bibliographystyle{utphys}
\bibliography{biblio}

\end{document}